\documentclass[referee]{aa} % for a referee version

\usepackage{natbib}
\bibpunct{(}{)}{;}{a}{}{,} % to follow the A&A style
\usepackage{graphicx}
\usepackage{lscape}
\usepackage{txfonts}
\usepackage{subfig}

\def\la{\mathrel{\mathchoice {\vcenter{\offinterlineskip\halign{\hfil
$\displaystyle##$\hfil\cr<\cr\sim\cr}}}
{\vcenter{\offinterlineskip\halign{\hfil$\textstyle##$\hfil\cr
<\cr\sim\cr}}}
{\vcenter{\offinterlineskip\halign{\hfil$\scriptstyle##$\hfil\cr
<\cr\sim\cr}}}
{\vcenter{\offinterlineskip\halign{\hfil$\scriptscriptstyle##$\hfil\cr
<\cr\sim\cr}}}}}

\def\p {$\pm$}

\def\kms {\hbox{${\rm km\, s}^{-1}$}} % km s-1 for LATEX
\def\cmsq  {$\hbox{{\rm cm}}^{-2}$}    %cm-2
\def\percc {$\hbox{{\rm cm}}^{-3}$}    %cm-3
\def\MOLH {\hbox{${\rm H}_2$}}  %H2
\def\MOLN {\hbox{${\rm N}_2$}}  % N2
\def\AMM {\hbox{${\rm NH}_{3}$}} %NH3
\def\HCOP {\hbox{${\rm HCO}^+$}}      %HCO+
\def\DCOP {\hbox{${\rm DCO}^+$}}    %DCO+
\def\HTHP {\hbox{${\rm H}_{3}^{+}$}}   %H3+
\def\HTDP {\hbox{${\rm H}_{2}{\rm D}^{+}$}}   %H2D+
\def\DTHP {\hbox{${\rm D}_{2}{\rm H}^{+}$}}   %D2H+
\def\DTHREEP {\hbox{${\rm D}_{3}^{+}$}}   %D3+
\def\HTHOP {\hbox{${\rm H}_{3}{\rm O}^{+}$}}  %H3O+
\def\NTHP {\hbox{${\rm N}_2{\rm H}^+$}} % N2H+
\def\NTDP {\hbox{${\rm N}_2{\rm D}^+$}} % N2D+
\def\trans {\hbox{${(1_{1,0}-1_{1,1}}$})} % o-H2D+ (1_1,0-1_1,1)

\begin{document}

\title{Survey of ortho-\HTDP (1$_{1,0}$--1$_{1,1}$) in dense cloud cores}
\author{P. Caselli\inst{1,2} \and C. Vastel\inst{3} \and C. Ceccarelli\inst{4} \and F.F.S. van der Tak\inst{5,6} 
       \and A. Crapsi\inst{7} \and A. Bacmann\inst{4,8}}
\institute{School of Physics and Astronomy, University of Leeds, Leeds LS2 9JT, UK  \email{p.caselli@leeds.ac.uk}
\and INAF--Osservatorio Astrofisico di Arcetri, Largo E. Fermi 5, 
I--50125 Firenze, Italy \and Centre d'Etude Spatiale des Rayonnements, CESR/CNRS-UPS, BP 4346, 
31028 Toulouse Cedex 04, France \and Laboratoire d'Astrophysique, Observatoire de Grenoble, BP 53, 38041 
Grenoble Cedex 9, France \and Max-Planck-Institut f\"ur Radioastronomie, Auf dem H\"ugel 69, 53121 Bonn, Germany 
\and  National Institute for Space Research 
(SRON), Postbus 800, 9700 AV Groningen, The Netherlands  \and Leiden Observatory, P.O. Box 9513, 2300 RA Leiden, The Netherlands 
\and Universit\'e Bordeaux 1, CNRS, OASU, UMR 5804, 33270 Floirac, France}    
\date{}    

  \abstract
  {}
   {We present a survey of the  ortho--\HTDP (1$_{1,0}$--1$_{1,1}$) line 
toward a sample of 10 starless cores and 6 protostellar cores, carried out at 
the Caltech Submillimeter Observatory. The high diagnostic power of this line 
is revealed for the study of the chemistry, and the evolutionary and dynamical status of 
low-mass dense cores.}
   {The derived ortho--\HTDP \ column densities ($N(ortho-\HTDP)$) are compared 
with predictions from simple chemical models of centrally concentrated cloud cores.} 
   {The line is detected in 7 starless
cores and in 4 protostellar cores.  $N(ortho-\HTDP)$ ranges between 2 and 
40$\times$10$^{12}$~cm$^{-2}$ in starless cores and between 2 and 
9$\times$10$^{12}$~cm$^{-2}$ in protostellar cores.
  The brightest lines are detected toward 
the densest and most centrally concentrated starless cores, where the CO depletion
factor and the deuterium fractionation are also largest. The 
large scatter observed in plots of $N(ortho-\HTDP)$ vs. the observed deuterium 
fractionation and vs. the CO depletion factor is likely to be due to variations 
in the ortho--to--para (o/p) ratio of \HTDP \ from $>$ 0.5 for $T_{\rm kin}$ $<$ 10~K gas in 
pre--stellar cores to $\simeq$ 0.03 (consistent with $T_{\rm kin}$ $\simeq$ 15~K for 
protostellar cores).  The two Ophiuchus cores in
our sample also require a relatively low o/p ratio ($\simeq$ 0.3). Other parameters, including the
cosmic-ray ionization rate, the CO depletion factor (or, more in general, the depletion
factor of neutral species), the volume density, the fraction of dust grains and PAHs 
also largely affect the ortho--\HTDP \ abundance.
In particular, gas temperatures above 15~K, low
CO depletion factors and large abundance of negatively charged small
dust grains or PAHs drastically reduce the deuterium fractionations to 
values inconsistent with
those observed toward pre--stellar and protostellar cores.
The most deuterated and \HTDP--rich objects (L~429, L~1544, L~694-2 and L~183) are reproduced by 
chemical models of centrally concentrated (central densties $\simeq$10$^{6}$ cm$^{-3}$) cores with 
chemical ages between 10$^4$ and 10$^6$~yr. Upper 
limits of the para-\HTHOP (1$_1^-$--2$_1^+$) and 
para--\DTHP (1$_{1,0}$--1$_{0,1}$) lines are also given. The upper limit to the 
 para-\HTHOP \ fractional abundance is $\simeq$10$^{-8}$ and we find 
an upper limit to the  para--\DTHP /ortho--\HTDP \ column density ratio equal to 
1, consistent with chemical model predictions of high density (2$\times$10$^6$ cm$^{-3}$) 
and low temperature ($T_{\rm kin}$ $<$ 10~K) clouds. }
   {Our results point out the need for better determinations of temperature and 
density profiles in dense cores as well as for observations of para-\HTDP .}

\maketitle
%
%________________________________________________________________

\section{Introduction}

  In the past decade, astrochemistry has become more 
and more crucial in understanding the structure and evolution of
star forming regions.  There are no doubts that stars like our Sun 
form in gas and dust condensations within molecular clouds and that the 
process of star formation can only be understood by means of detailed 
observations of the dust (good probe of the most abundant and elusive 
molecule, \MOLH) and molecular lines (unique tools to study kinematics
and the chemical composition).  

Millimeter and submillimeter continuum 
dust emission observations \citep[see][for a detailed review of this topic]{wac07} 
have provided a very good probe of the density 
structure of dense cores, although uncertainties are still present 
regarding the dust opacity and temperature, both likely to change 
within centrally concentrated objects \citep[but such variations have so far been 
hard to quantify observationally, see e.g.][]{bga03,plb03,pbm04}.
%(but such variations have so far been 
%hard to quantify observationally; see e.g. \cite{bga03}; \cite{pbm04}; 
%\cite{plb03}).  
Stellar counts in the near-infrared provide an alternative
way of measuring the dust (and \MOLH ) column and cloud structure 
\citep{llc94}, independent of any variation in dust properties, but they 
cannot probe regions with extinctions above about 40-50 mag \citep{all98}, 
i.e. the central zone of very dense cores, such as L~1544, where 
$A_{\rm V} \simeq$ 100 mag within 11$^{\arcsec}$ 
\citep{wma99}. 
 It appears that many starless cores can be 
approximated as Bonnor--Ebert spheres \citep{e55,b56}, 
with values of the central densities ranging from about 10$^5$ \percc \
\citep[as in the case of B~68;][]{all01} to 10$^6$ \percc \ 
(for e.g., L~1544, L~183, and L~694-2; \citealt{wma99}, \citealt{plb03}, \citealt{hwl03}). 
At the lower end of the central density 
range, dense cores appear to be isothermal, with gas temperatures
close to 10~K \citep{gwg02,tmc04}, whereas higher density cores 
have clear evidences of temperature drops in the central few thousand 
AU, with dust temperatures approaching about 7~K 
\citep[][see also Bergin \& Tafalla 2007 for a comprehensive review on starless cores]{ers01,plb03,pbm04,sg05,pbc07,ccw07}

\citet{kf05} have proposed that the 
"shallower" cores (such as B~68) are in approximate equilibrium and 
will not evolve to form protostars, whereas the centrally concentrated ones
(such as L~1544) are unstable cores that are proceeding toward 
gravitational collapse and the formation of protostars. Indeed, this is 
in agreement with the findings of \cite{lba02}, who claim that 
B~68 is oscillating around an equilibrium state, and those of \citet{cwz02a} and 
\citet{vcc05}, who studied the kinematic 
structure of L~1544 and found that it is consistent with contraction in the 
core nucleus (or central contraction). 
 
To trace the gas properties, \AMM \ and \NTHP \ have been 
extensively used for several years \citep{bm89} and they 
seem to trace quite similar conditions, having comparable morphologies and 
line widths \citep{bcm98,cbm02,tmc02,tmc04}, 
despite of the (two orders of magnitude) difference in the 
 critical densities of the most frequently 
observed transitions (\AMM (1,1) and \NTHP (1-0), see \citealt{pbc07}).  
This is quite lucky 
for astronomers considering that only relatively recently it has been 
realized that these two species are among the few that are left in 
the gas phase at volume densities above $\sim$10$^5$ \percc . In fact, 
CO, CS and, in general, all the carbon bearing species so far observed 
\citep[with the exception of CN;][]{hwp08} 
are heavily affected by freeze-out in the central parts of dense starless cores 
\citep{klv96,wlv98,cwt99,bcl01,blc02,bah02,cwz02b,tmc02,blc03,ccw04,tmc04,ppa05,ccw05,tsm06,pbc07}. 
The freeze-out of neutral species \citep[in particular of CO,][]{dl84,rm00a,rm00b}, boosts the deuterium fractionation in 
species such as \NTHP , \AMM , H$_2$CO , and \HCOP \citep[see e.g.][]{bll95,trf00,cwz02b,blc03,ccw05,lgr06,glp06}.
Also in star forming cores, in particular in the direction of 
Class 0 sources, the first protostellar stage \citep[e.g.][]{awb00}, right after the pre-stellar phase, 
the deuterium fractionation is found to be very large \citep{ccl98,lcc02,lgp02,lrg02,vsm02,pct02,vpc03,pch04,ccw04,mcr05,pct06,cch07}.  
This is thought to be the signature of a (recent) past in which the star forming cloud core
experienced the low temperature and high density conditions typical of 
the most centrally concentrated starless cores, where some 
deuterated molecules (e.g. deuterated ammonia) are formed in the 
gas phase (and stored on dust surfaces) whereas others (such as deuterated
methanol and formaldehyde) are likely formed onto dust surfaces and then 
released to the gas phase partially upon formation \citep{gpc06,gwh07} 
as well as via the interaction with the newly born protostar, which can (i) heat 
dust grains, leading to mantle evaporation \citep[as in Hot Cores and Corinos;][]{t90,ctc03,bcl04,bcn04,bcw07}, and/or (ii) 
"erode" dust mantles via sputtering in shocks produced by the associated energetic outflows \citep[e.g.][]{lgp02}.  

Questions that are still open are: (1) are \NTHP , \AMM \ and their deuterated
forms really tracing the inner portions of centrally concentrated cores 
on the verge of star formation?  Although Bergin et al. (2002), Pagani 
et al. (2005, 2007) found evidences of depletion in the center of B~68 and 
L~183, respectively, there are no signs of freeze--out for NH$_3$ in L~1544
(Crapsi et al. 2007) and for both nitrogen bearing species in L~1517B and L~1498 
(Tafalla et al. 2004). At densities above $\sim$10$^6$ \percc ,
the freeze--out time scale is quite short ($\sim$1,000 yr) and all heavy
species are expected to condense onto grain mantles. Moreover, recent 
laboratory measurements clearly show that \MOLN , the parent species 
of both \AMM \ and \NTHP , should freeze--out at the same rate as 
CO \citep[having similar binding energies and sticking probabilities;][]{ovf05,bfo06}. (2) For how long the high 
degree of deuterium fractionation observed in starless cores is maintained 
after the formation of a protostellar object? 

The detection of strong ortho-\HTDP \trans \ emission in the direction of L~1544
\citep{cvc03}, and the conclusion that \HTHP , with its deuterated counterparts, is one of the most
abundant molecular ions in core centers, have opened a new way to study the 
chemical evolution \citep{rhm03,rhm04,wfp04,fpw04,fpw05,fpw06a,fpw06b,ahr05} and the kinematics 
\citep{vcc05} of the central few thousand AU of starless cores.  
Thus, \HTDP \  is an important tool to understand the chemical and physical properties 
of the material out of which protoplanetary disks and ultimately planetary systems form. 

In the gas phase at temperatures below $\sim$20~K, the deuterium fractionation is mostly regulated by the 
proton--deuteron exchange reaction:
\begin{eqnarray}
\HTHP + {\rm HD} \rightarrow \HTDP + \MOLH + \Delta {\rm E} ,
\label{e_htdp}
\end{eqnarray}
where $\Delta$E \citep[= 230~K,] []{mbh89} prevents the reverse 
reaction to be fast in cold regions, unless a significant fraction of 
\MOLH \ is in ortho form \citep{ghr02,wfp04}. 
\citet{fpw06a} showed that values of the ortho--to--para (o/p) \MOLH \ ratio
much higher than 0.03 are inconsistent with the observed high levels of 
deuteration of the gas (see their Fig.~6). 
The above reaction, together with the freeze--out of neutral species
\citep[which boosts up the production rate of \HTDP \ compared to \HTHP ; e.g.][]{aoi01}, 
allows the \HTDP /\HTHP \ ratio to overcome the cosmic D/H ratio by several orders of magnitude. 
In fact, not only ortho--\HTDP \trans \ has been found to be $\simeq$1~K strong in L~1544, but
also \DTHP \ has been detected toward another starless condensation 
\citep[16293E, in Ophiuchus;][]{vpy04}.  The ortho--\HTDP \trans \
line has also recently been mapped in L~1544 \citep{vcc06}, finding
that the \HTDP \ emitting region has a radius of about 5000 AU, comparable
to the size of the \NTDP (2-1) map made in the same region by \citet{cwz02a}. 
%Thus, \HTDP \ is a very good tracer of the denser regions
%of cores and, being a single line (i.e. no resolvable hyperfine structure
%is present, unlike for the \NTHP \ and \NTDP \ lines, where hyperfine component 
%blending in the high--J transitions prevents an accurate determination of the
%velocity field in the high density core regions) can help in studying the kinematics.

In this paper we present new ortho--\HTDP \trans \ observations, carried 
out at the Caltech Submillimeter Observatory (CSO) antenna, in the direction
of 10 starless cores and 6 cores associated with very young protostellar 
objects.  As it will be shown, the line has been detected in 7 of the 10 
starless cores and in 4 out of 6 star forming cores. In Sect.~\ref{sobs} the
observational details are given. Results and ortho--\HTDP \ spectra 
are shown in Sect.~\ref{sres}, together with a brief discussion on the 
upper limits of the para-\DTHP (1$_{1,0}$-1$_{0,1}$) lines (para-\HTHOP (1$_1^-$--2$_1^+$) 
upper limits can be found in the on-line appendix). \HTDP \ column densities are derived 
in Sect.~\ref{sana}.  A chemical discussion, aimed at interpreting 
the observations, is described in Sect.~\ref{schem} and conclusions
are in Sect.~\ref{scon}.  

\section{Observations}
\label{sobs}

\subsection{Technical details}
\label{stechnical}

Observations of the ortho--\HTDP \trans \ line ($\nu_0$ = 372.421385(10) GHz; 
\citealt{ah05}) were carried out at the Caltech Submillimeter Observatory
(CSO) on Mauna Kea (Hawaii), between October 2002 and April 2005. The spectra
were taken in wobbler switching mode, with a chop throw of 300\arcsec .
The backend used was an acousto-optical spectrometer (AOS) with 50 MHz bandwidth. 
The velocity resolution, as measured from a frequency comb scan, is 0.1 \kms .
The beam efficiency ($\eta_b$) at $\nu$ = 372 GHz was measured on Saturn, Mars and 
Jupiter and is listed in Tab.~\ref{tfit}. Measurements for extended sources were made 
for only few sources (L~1544, L~183, NGC~1333--DCO$^+$, B~1, NGC~2264G-VLA2) 
and found to be $\sim$ 70\%, compared to 60\% on planets measurements. \HTDP 
is likely to be extended (e.g. L~1544: \citealt{vcc06}; L~183:  \citealt{vpc06}). 
Consequently, we used the extended source beam efficiency whenever available. However note that 
the difference cannot be much significant. At 372 GHz, the
CSO 10.4-m antenna has a half power beamwidth of about 22$^{\arcsec}$.

Similar setups were used for the para--\HTHOP (1$_1^-$-2$_1^+$) line at 
307.1924100 GHz (JPL catalogue to be found at http://spec.jpl.nasa.gov/), 
which was observed at CSO in October 2002 and June 2003.   
Tab.~\ref{thtop} present the beam efficiencies that were used for H$_3$O$^+$ data. 
Pointing was measured every two hours and found to be better than 3\arcsec . 
%\begin{table}
%\begin{center}
%\caption{Observational Parameters \label{tobs}}
%\begin{tabular}{lccccc}
%\hline\hline
%Period &  225 GHz  & $T_{\rm sys}$(\HTDP) & $\eta_{B}$(\HTDP ) &
% $T_{\rm sys}$(\HTHOP) & $\eta_{B}$(\HTHOP )  \\
%       &  opacity & (K) & & (K) &  \\
%\hline
%October 2002  & 0.05 & 1200 & 0.60 & 300 & 0.70 \\
%June 2003     & 0.09 & 2000 & 0.30 & 400 & 0.50 \\
%December 2003 & 0.05 & 1300 & 0.50 & ... & ...  \\
%February 2004 & 0.05 & 1300 & 0.70 & ... & ...  \\
%November 2004 & 0.04 & 1200 & 0.70 & ... & ...  \\
%December 2004 & 0.04 & 1200 & 0.70 & ... & ...  \\
%April 2005    & 0.03 & 1300 & 0.70 & ... & ...  \\
%\hline
%\end{tabular}
%\end{center}
%\end{table}

Observations of the para-D$_2$H$^+$ (1$_{1,0}$-1$_{0,1}$) line \citep[$\nu$ =
691.660483(20) GHz;][]{ah05} were carried out at CSO between
April 2003 and April 2005 under very good weather conditions (225 GHz zenith
opacity less than 0.065). We used the 50 MHz AOS with a spectral
resolution better than 0.04 km~s$^{-1}$. The
observations were performed using the wobbler with a chop throw between
150$\arcsec$ and 180$\arcsec$ according to its stability. The beam
efficiency was carefully and regularly checked on Mars, Venus, Saturn and
Jupiter, and found to be $\sim$ 40\%. For more extended sources, the beam 
efficiency has been measured to be $\sim$ 60\%. This value has been adopted 
for 16293E, the only source where D$_2$H$^+$ has been detected, assuming that 
in this case the emission covers an area not significantly smaller than the 
beam. For the other sources, we use $\eta_b$ = 40\% (Tab.~\ref{tdthp}) and 
consider a factor of 1.5 uncertainty in the column density upper limit value 
due to the unknown source size.  Pointing was monitored every 1.5
hours and found to be better than 3$^{\arcsec}$. At 692 GHz, the
CSO 10.4-m antenna has a half power beamwidth of about 11$^{\arcsec}$.

\subsection{Source Selection}

The source list is in Tab.~\ref{tsample}, which reports the coordinates, 
the Local Standard of Rest velocity ($V_{\rm LSR}$) at which we centered
our spectra, and the distance to the source. 

\begin{table}
\begin{center}
\caption{Source sample \label{tsample}}
\begin{tabular}{lcccc}
\hline\hline
Source & RA(J2000) & Dec(J2000) & $V_{\rm Lsr}$ & Distance \\
Name   & [h m s] & [$\deg$ \arcmin\ \arcsec] & (\kms ) & (pc) \\
\hline
\multicolumn{5}{c}{Starless Cores} \\
\hline
L~1498 & 04 10 53.67 & +25 10 18.12 & 7.80 & 140  \\
TMC--2 & 04 32 44.03 & +24 23 32.56 & 6.10 & 140 \\
TMC--1C & 04 41 38.81 & +26 00 21.98 & 5.20 & 140 \\
L~1517B & 04 55 18.00 & +30 37 43.84 & 5.80 & 140 \\
L~1544  & 05 04 17.23 & +25 10 42.70 & 7.14 & 140 \\
L~183   & 15 54 08.56 & -02 52 48.99 & 2.50 & 110 \\
Oph~D  & 16 28 28.56 & -24 19 25.03 & 3.40 & 165 \\
B~68 & 17 22 38.64 & -23 49 46.03 & 3.40 & 125 \\
L~429   & 18 17 05.53 & -08 13 29.94 & 6.82 & 200 \\
L~694-2 & 19 41 05.03 & +10 57 01.99 & 9.60 & 250 \\
\hline
\multicolumn{5}{c}{Protostellar Cores} \\
\hline
NGC 1333 DCO$^+$ & 03 29 12.10 & +31 13 26.30 & 6.00 & 350  \\
B~1        & 03 33 20.84 & +31 07 34.27 & 6.00 & 350  \\
IRAM~04191 & 04 21 56.91 & +15 29 46.10 & 6.60 & 140  \\ 
L~1521F$^a$ & 04 28 39.80 & +26 51 35.00 & 6.48 & 140 \\
Ori~B9 & 05 43 08.17 & -01 15 11.76 & 9.10 & 450 \\
NGC~2264G--VLA2 & 06 41 11.09 & +09 55 59.01 & 8.00 & 800 \\ 
\hline
\end{tabular}
\begin{list}{}{}
\item[$^{\mathrm{a}}$] Embedded low luminosity object recently detected by Spitzer \citep{bme06}.
\end{list}
\end{center}
\end{table}

The selection criteria for starless cores is similar to that described 
in \citet{ccw05}, where sources with bright continuum and 
 \NTHP \ emission have been selected to include chemically evolved cores, 
where CO is significantly frozen onto dust grains and where 
\HTDP \ is thus expected to be more abundant.  The sample consists 
of "shallow" cores, with central densities of $\sim$10$^5$ \percc \ (L~1498, 
TMC--2, L~1517B, B~68) and more centrally concentrated ones, with central
densities of $\sim$10$^6$ \percc \ (TMC--1C, L~1544, L~183, Oph~D, L~429 and
L~694-2).  

The star forming regions observed have been selected as being representative
of the early phases of protostellar evolution, so that any detection of 
\HTDP \ will be interesting to compare with starless cores and see if 
any evolutionary trend can be highlighted. Among the protostellar cores we
selected: \\
(1) NGC 1333 \DCOP , close to IRAS 4A, in Perseus, where large 
abundances of deuterated species have been observed (ND$_3$, \citealt{vsm02}; 
NH$_2$D, \citealt{h03}; D$_2$S, \citealt{vpc03}; 
ND$_2$H, \citealt{rlv05}). \\
(2) B~1 is one of the highest column 
density cores in the Perseus Complex, with active low--mass star formation
going on \citep[e.g.][]{hkm99} and with large deuterium fractionations,
as shown by the detection of triply deuterated ammonia \citep[ND$_3$,][]{lrg02,lgr06}, 
doubly deuterated hydrogen sulfide \citep[D$_2$S,][]{vpc03}, 
and doubly deuterated thioformaldehyde \citep[D$_2$CS,][]{mcr05}.   \\
(3) IRAM 04191, a very low luminosity Class 0 source in Taurus, 
driving a powerful outflow, but embedded in a dense core which appears to 
maintain many of the starless core characteristics \citep{ba04}. \\
(4) L~1521F, initially selected as a starless core with chemical and physical
structure similar to L~1544 \citep[but different kinematics;][]{ccw04}, 
and recently found associated with a $L <$ 0.07 L$_{\odot}$ protostellar 
object thanks to the high sensitivity of the Spitzer Space Telescope 
\citep{bme06}. \\
(5) Ori~B9, a massive dense core in Orion B, 
with peculiarly narrow molecular line widths and low gas temperature 
\citep{lbs91,hww93,cm95}, thus an ideal target (among massive cores) to detect deuterated species. \\
(6) NGC2264--VLA2, studied by, e.g., \citet{wec95}, 
who found the Class 0 driving source of the bipolar outflows, 
\citet{geh00}, who determined the gas temperature and volume density of 
the surrounding core, and by \citet{lcc02}, who observed 
D$_2$CO and found an extremely large deuterium fractionation 
([D$_2$CO]/[H$_2$CO] = 0.4, equivalent to an enrichment over the
cosmic D/H  ratio of more than 9 orders of magnitude). 

\section{Results}
\label{sres}

\subsection{ortho-\HTDP}

\begin{figure}
\centering
\includegraphics[width=0.45\textwidth]{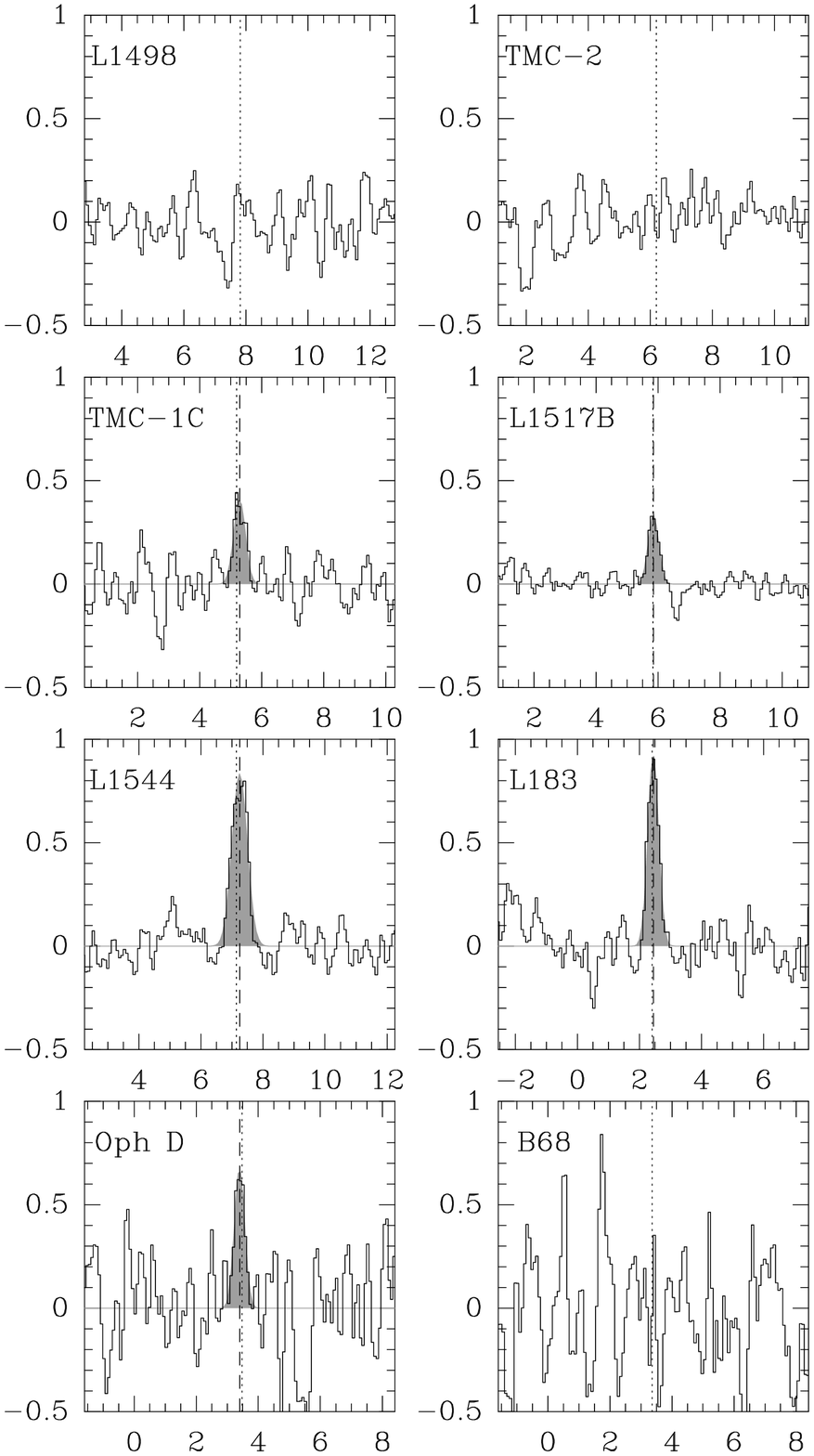}
\quad
\includegraphics[width=0.45\textwidth]{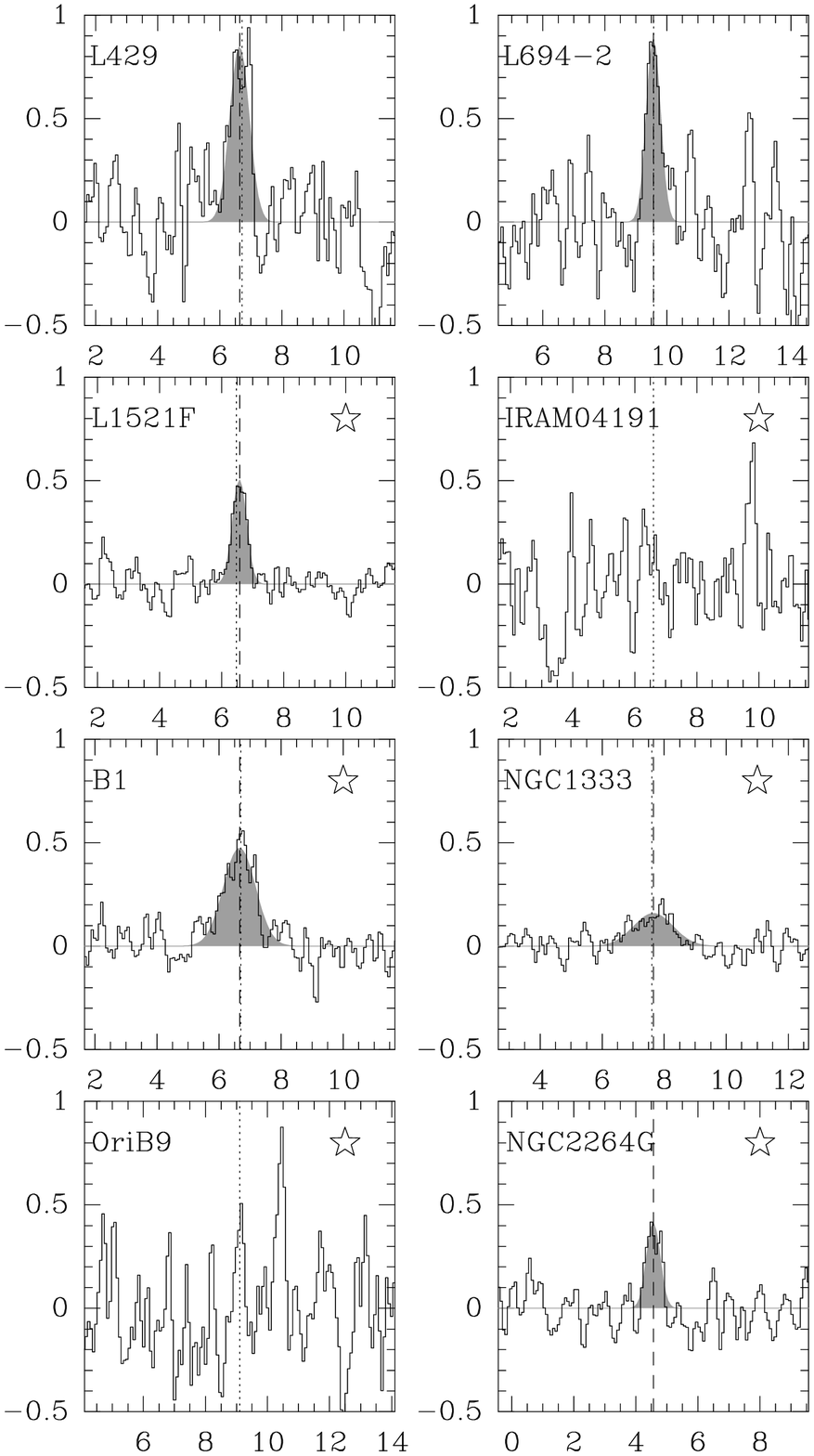}
\caption{ortho--\HTDP \trans \ spectra toward the 16 sources of our sample.
The units are main beam brightness temperature in K ($y$-axis, assuming a
unity filling factor) and velocity in \kms \ ($x$-axis). A star in the top right indicates 
dense cores associated with protostellar objects.
 The vertical dotted line is the
$V_{\rm LSR}$ velocity measured with \NTHP (1-0), whereas the vertical
dashed line marks $V_{\rm LSR}$ as measured with the present \HTDP \
observations.  Within the uncertainties, the two values are identical.
Note that the strongest emission is
present in the densest starless cores (L~1544, L~183, L~429, L~694-2) and in
the star forming regions L~1521F, B~1, and NGC~2264G.  Note also the
large variation in linewidth among the various sources.}
\label{fspectra}
\end{figure}

The ortho--\HTDP \trans \ spectra are shown in Fig.~\ref{fspectra}, 
whereas the results of Gaussian fits to the lines are listed in 
Tab.~\ref{tfit}. One striking thing which stands out from the figure and 
the table is the large variation in intensity (factor of 5) and linewidths
(factor of 4), and not just between starless cores and protostellar cores.  
In column 5 of Tab.~\ref{tfit} we also report the non-thermal line width,
defined as (see \citealt{mlf91}):
\begin{eqnarray}
\Delta {\rm v}_{\rm NT} & = & \sqrt{\Delta {\rm v}_{\rm obs}^2 - \Delta {\rm v}_{\rm T}^2},
\end{eqnarray}
where $\Delta {\rm v}_{\rm obs}$ ($\equiv \Delta {\rm v}$ in Tab.~\ref{tfit}) is the 
observed line width and $\Delta {\rm v}_{\rm T}$ is the thermal linewidth of the 
observed molecule, 
calculated assuming the kinetic temperatures listed in column 3 of 
Tab.~\ref{tcolumn}. It is interesting to note that the ortho--\HTDP \trans \
non-thermal line width is on average two times larger than that derived 
from the \NTDP (2-1) data of \citet{ccw05} (using the same kinetic temperature),
the only exception being B~68, where both the \NTDP (2-1) and 
ortho--\HTDP \trans \ lines are totally thermally broadened. The larger 
non-thermal line widths (and the moderate optical depths, see Sect.~\ref{sana}) 
indicate that the ortho--\HTDP \ lines 
are tracing a region within the cores with more prominent internal motions than 
the \NTDP (2--1) lines, even in L~1544, where the emission of these two lines 
has the same morphology and extension \citep{vcc06}. 

\begin{table}
\begin{center}
\caption{Gaussian fits of the ortho--\HTDP \trans \ lines in the source
sample. \label{tfit}}
\begin{tabular}{lccccccc}
\hline\hline
Source & $\int T_{mb}dV$ & $V_{\rm LSR}$ & $\Delta {\rm v}$ &
        $\Delta {\rm v}_{\rm NT}$$^{\mathrm{a}}$ &
        $T_{\rm mb}$ & $\eta_{b}$  &  $T_{\rm rms}$  \\
Name   & (K \kms) & (\kms ) & (\kms ) & (\kms ) &(K) &   &  (K) \\
\hline
\multicolumn{8}{c}{Starless Cores} \\
\hline
L~1498            & ...         & ...         & ...         & ...  & ... & 0.6  &  0.13 \\
TMC--2           & ...         & ...         & ...         & ...  & ... & 0.6  &  0.13 \\
TMC--1C          & 0.19\p 0.02 & 5.29\p 0.03 & 0.44\p 0.06 & 0.34\p 0.09 &
        0.41 & 0.5  &  0.12 \\
L~1517B           & 0.14\p 0.01 & 5.85\p 0.02 & 0.40\p 0.04 & 0.29\p 0.05 &
        0.33 & 0.5  &  0.06 \\
L~1544$^{\mathrm{b}}$            & 0.50\p 0.03 & 7.28\p 0.02 & 0.50\p 0.03 & 0.41\p 0.03 &
        0.93 & 0.7  &  0.11 \\
L~183             & 0.40\p 0.02 & 2.45\p 0.01 & 0.41\p 0.03 & 0.30\p 0.04 &
        0.92 & 0.7  &  0.12 \\
Oph~D           & 0.28\p 0.05 & 3.40\p 0.03 & 0.39\p 0.09 & 0.27\p 0.11 &
        0.67 & 0.4 &  0.26 \\
B~68              & ...         & ...         & ...         & ...  & ... & 0.4  &0.32 \\
B~68$^{\mathrm{c}}$ & 0.078\p 0.015 & 3.36\p 0.05 & 0.33\p 0.10 & 0.0\p 0.1 &
        0.22 &   &  0.08 \\
L~429             & 0.65\p 0.05 & 6.64\p 0.03 & 0.73\p 0.06 & 0.67\p 0.07 &
        0.85 & 0.4  &  0.21 \\
L~694-2          & 0.50\p 0.05 & 9.56\p 0.03 & 0.53\p 0.07 & 0.45\p 0.08 &
        0.88 & 0.45  &  0.24 \\
\hline
\multicolumn{7}{c}{Protostellar Cores} \\
\hline
NGC 1333 DCO$^+$ & 0.25\p 0.02 & 7.65\p 0.06 & 1.49\p 0.12 & 1.44\p 0.12 &
        0.16 & 0.7  &  0.05 \\
B~1               & 0.60\p 0.03 & 6.66\p 0.03 & 1.20\p 0.08 & 1.14\p 0.08 &
        0.47 & 0.7  &  0.09 \\
IRAM~04191        & ...         & ...         & ...         & ...  & ... & 0.65  &  0.23 \\
L~1521F           & 0.28\p 0.02 & 6.58\p 0.02 & 0.53\p 0.03 & 0.41\p 0.04 &
        0.50 & 0.6  &  0.07 \\
Ori~B9            & ...         & ...         & ...         & ...  & ... & 0.5  &  0.25 \\
Ori~B9$^{\mathrm{d}}$ & 0.07\p 0.03 & 9.20\p 0.06 & 0.42\p 0.15 &
        0.25\p 0.25 & 0.16 & &  0.04 \\
NGC~2264G--VLA2   & 0.22\p 0.02 & 4.56\p 0.03 & 0.51\p 0.04 & 0.29\p 0.07 &
        0.41 & 0.7  &  0.11 \\
\hline
\end{tabular}
\begin{list}{}{}
\item[$^{\mathrm{a}}$] Non-thermal linewidth, assuming the gas kinetic temperature listed in Tab.~\ref{tcolumn} (see text).
\item[$^{\mathrm{b}}$] Data from \citet{vcc06}.
\item[$^{\mathrm{c}}$] Data from \citet{hce06}. The thermal linewidth at 10~K is 0.34 \kms.
\item[$^{\mathrm{d}}$] Data from \citet{hhl06}.
\end{list}
\end{center}
\end{table}

%\begin{landscape}
\begin{table*}
\caption{ortho--\HTDP column densities. \label{tcolumn}}
\centering
\begin{tabular}{lcccccccc}
\hline\hline
Source & $n(\MOLH )$ & $T_{\rm kin}$ & Ref.$^{\mathrm{a}}$ & $N(\MOLH )$$^{\mathrm{b}}$ & Ref. &
         $T_{\rm ex}$$^{\mathrm{c}}$ & $\tau$$^{\mathrm{c}}$ &
         $N$(ortho--\HTDP )$^{\mathrm{c}}$ \\
Name   & (\percc ) & (K) &  & (10$^{22}$ \cmsq) & & (K) &  &
         (10$^{13}$ \cmsq ) \\
\hline
\multicolumn{9}{c}{Starless Cores} \\
\hline
L~1498            & 1$\times$10$^5$ & 10    & 4   & 3.2 & 7
                 & 7.5,5.5 & $<$0.3,$<$0.9 & $<$0.7,$<$2 \\
TMC--2           & 3$\times$10$^5$ & 10$^{\mathrm{d}}$ & 7 & 6.0 &
                   7 & 8.7,5.9 & $<$0.2,$<$0.6 & $<$0.5,$<$1.5 \\
TMC--1C          & 1$\times$10$^6$ & 7 & 8,9 & 8.5 & 9 &
                   6.8,6.0 & 0.4,0.6 & 0.9,1.4\\
L~1517B           & 2$\times$10$^5$ & 9.5 & 4 & 3.7 & 7 &
                   8.0,5.6 & 0.2,0.9 & 0.4,1.8 \\
L~1544            & 2$\times$10$^6$ & 7 & 21  & 13 & 7$^{\mathrm{i}}$ &
                   6.9,6.5 & 1.2,1.7 & 3.2,4.5 \\
L~183             & 2$\times$10$^6$ & 7     & 12,22 & 14 & 12 &
                   6.9,6.5 & 1.1,1.6 & 2.5,3.5 \\
Oph~D           & 5$\times$10$^5$ & 7 & 23 & 11 &
                   23 & 6.7,5.8 & 0.8,2.0 & 1.6,4.0 \\
B~68$^{\mathrm{e}}$  & 3$\times$10$^5$ & 10 & 24,13 & 1.4 & 7 &
                   8.7,5.9 & 0.1,0.3 & 0.2,0.5 \\
L~429             & 6$\times$10$^5$ & 7$^{\mathrm{d}}$ & 7 & 12 &
                   7$^{\mathrm{i}}$ & 6.8,6.3 & 1.1,1.8 & 4.1,6.7 \\
L~694-2          & 9$\times$10$^5$ & 7$^{\mathrm{d}}$ & 7 & 11 &
                   7$^{\mathrm{i}}$ & 6.9,6.3 & 1.1,2.0 & 3.2,5.7 \\
16293E$^{\mathrm{f}}$ & 2$\times$10$^6$ & 12 & 19 & 50 & 19 &
                   12,10 & 0.5,0.7 & 1.2,1.7 \\
\hline
\multicolumn{9}{c}{Protostellar Cores} \\
\hline
NGC 1333 DCO$^+$ & 1$\times$10$^6$$^{\mathrm{d}}$       & 13    & 1   & 31 & 14 &
                   12,8.7 & 0.03,0.1 & 0.35,0.59 \\
B~1               & 3$\times$10$^6$ & 12    & 2,3 & 21 & 15 &
                   12,10 & 0.1,0.1 & 0.9,1.1 \\
IRAM~04191        & 5$\times$10$^5$ & 10    & 5   & 17 & 16 &
                   9.2,6.8 & $<0.3,<0.6$ & $<0.8,<1.5$ \\
L~1521F           & 1$\times$10$^6$ & 9.3  & 6,20  & 14 &
                   7 & 8.9,7.3 & 0.2,0.4 & 0.7,1.1 \\
Ori~B9$^{\mathrm{g}}$ & 7$\times$10$^4$ & 10 & 10 & 4.0 & 10 &
                   6.8,4.2 & 0.1,0.4 & 0.2,0.9 \\
NGC~2264G--VLA2   & 8$\times$10$^5$ & 15    & 11  & 10 & 17 &
                   14,9.2 & 0.1,0.2 & 0.3,0.5 \\
NGC 1333 IRAS 4A$^{\mathrm{h}}$ & 2$\times$10$^6$ & 13 & 18 & 31 & 18 &
                   13,10 & 0.03,0.04 & 0.2,0.3 \\
\hline
\end{tabular}
\begin{list}{}{}
\item[$^{\mathrm{a}}$] When more than one reference is listed, the first number refers to the  
source paper for the volume density and the second number refers to the source 
paper for the kinetic temperature. In the case of L~183,  both parameters can be found in 
the two listed papers.
\item[$^{\mathrm{b}}$] Values of $N(\MOLH )$ have been estimated assuming a constant gas temperature of 10~K along the line of sight, except for those objects with central temperatures 
of 7~K (see item "i" below).  
\item[$^{\mathrm{c}}$] The first value is calculated assuming a critical density $n_{\rm cr}$=10$^5$ \percc ,
whereas the second value is for $n_{\rm cr}$ = 10$^6$ \percc .
\item[$^{\mathrm{d}}$] Assumed value.
\item[$^{\mathrm{e}}$] ortho-\HTDP \ data from \citet{hce06}, but method described in the text.
\item[$^{\mathrm{f}}$] ortho-\HTDP data from \citet{vpy04}, but method described in the text.
\item[$^{\mathrm{g}}$] Estimates of $T_{\rm ex}$, $\tau$ and $N({\rm ortho-\HTDP} )$ based
on APEX data from \citet{hhl06},  but using the method described in the text.
\item[$^{\mathrm{h}}$] Data from \citet{svv99}, but method described in the text.
\item[$^{\mathrm{i}}$] The value of $N(\MOLH )$ given by \citet{ccw05} has been modified to take into account the temperature gradient (see Sect.~\ref{scolumn} for details). 
\end{list}
{References: (1) \citet{h03}; (2) \citet{lgr06};
(3) \citet{mcr05}; (4) \citet{tmc04}; (5) \citet{ba04}; (6) \citet{ccw04};
(7) \citet{ccw05}, in a 11\arcsec \ beam; (8) \citet{sg05}; (9) \citet{skg07}; 
(10) \citet{hww93}, in a 40\arcsec \ beam; (11) \citet{geh00}; 
(12) \citet{pbm04}; (13) \citet{hhj02};
(14) \citet{vsm02}, in a 13\arcsec \
beam; (15) \citet{hrf05}, in a 14\arcsec \ beam; (16) \citet{bad02},
in a 11\arcsec \ beam; (17) \citet{gct94}, within 20\arcsec ;
(18) \citet{svv99} ($N(\MOLH )$ measured in a 13\arcsec \ beam);
(19) \citet{vpy04} ($N(\MOLH )$ measured in a 13\arcsec \ beam
using \citealt{lgp02} data); (20) \citet{kwa07}; (21) \citet{ccw07};
(22) \citet{pbc07}; (23) \citet{sww07}; (24) \citet{bmv06}.}
\end{table*}
%\end{landscape}

In the category of "shallow" cores, 
we have 3 non detections among 4 objects. The only 
shallow core detected in our survey is L~1517B, which is in fact the most 
compact and centrally concentrated of its class and has one of the 
narrowest \HTDP \ lines observed so far (together with the denser core 
Oph~D): 0.4 \kms , only 1.2 times larger than the \HTDP \ thermal 
linewidth at 10 K, as measured with \AMM \ observations \citep{tmc04}. 
Recent observations carried out with the APEX telescope 
have actually revealed a probable ortho-\HTDP \ emission in one of our non--detected
shallow cores, B~68 \citep{hce06}. The detected line is indeed 
relatively faint ($T_{\rm mb}$ = 0.2 K), very narrow (0.3 \kms , 
practically thermal), and consistent with our 
non--detection ($T_{\rm rms}$ = 0.3 K, see Tab.~\ref{tfit}). The 
observations of the other two objects in this group
(L~1498, TMC--2) have better sensitivities 
($T_{\rm rms}$ $\simeq$ 0.13 K), but they may still hide the \HTDP \ line 
if its intensity is similar to the one in B~68. 

Among the five most centrally concentrated objects in the starless core 
sample, four show strong ($T_{\rm mb}$ = 0.7 -- 0.9 K) \HTDP \ emission, 
whereas TMC--1C has $T_{\rm mb}$ $\simeq$ 0.4 K.  The line widths span 
the range between 0.4 \kms \ (for Oph~D) and 0.7 \kms \ (for L~429), 
maybe reflecting 
contraction motions in different stages of core evolution 
or large optical depths (although 
simple analysis seems to discard this last hypothesis; see Sect.~\ref{sana}). 

Also in the six (young) protostellar cores, the detection rate is quite 
large, with four \HTDP \ lines detected. L~1521F has a line shape quite similar
to that in L~1544 ($\Delta {\rm v}$ = 0.5 \kms ), but the brightness temperature is 
1.7 times lower, in agreement with the two times lower deuterium 
fractionation observed \citep[$N(\NTDP )/N(\NTHP )$ = 0.1 and 0.2, in L~1521F and 
L~1544, respectively;][]{ccw04,cwz02a}. B~1 and 
NGC~1333--\DCOP \ present the largest linewidth in the sample, suggesting
that the active star formation is probably injecting energy in form of 
non--thermal motions and turbulence. We are confident that the broad line in 
NGC~1333--\DCOP \ is not a baseline artifact, in particular because both the centroid
velocity and the linewidth are coincident (within the errors) with those 
observed in D$_2$S by \citet{vpc03} (but not with the ND$_3$ line 
observed by \citealt{vsm02}, which remains a puzzle).  The \HTDP \ 
line in NGC~2264G is narrower than in B~1 and NGC~1333--\DCOP \ and more similar 
to L~1521F, probably indicating that the circumstellar environment is still
quite pristine. 

\subsection{para-\DTHP }

The para-\DTHP (1$_{1,0}$-1$_{0,1}$) line has been searched for in four
sources (B~1, NGC~1333--\DCOP , NGC~2264G--VLA2, and L~183).  Tab.~\ref{tdthp}
lists the spectral resolution ($\Delta {\rm v}_{\rm res}$, column 2), the system
temperature ($T_{\rm sys}$, column 3) and the integration time ($t_{\rm int}$,
column 4) of the observations.  The rms noise and the upper limits of the
radiation temperature
(or brightness temperature, assuming a unity filling factor) are in columns
5 and 6, respectively.  From these data, we calculated the corresponding
upper limits of the para-\DTHP \ column density in each source,
applying the same method as for ortho-\HTDP \ (see Sect.~\ref{sana})
and assuming a critical density for the transition of 10$^5$~\percc , a line width
equal to that measured for the \HTDP \ line (Tab.~\ref{tfit}), except for 
16293E (for which we used the para-\DTHP \ line width observed by \citealt{vpc03}), 
the kinetic temperature and \MOLH \ volume densities listed in Tab.~\ref{tcolumn},
and the following parameters for the para-\DTHP (1$_{1,0}$-1$_{0,1}$)
transition: frequency $\nu_0$ = 691.660483 GHz, Einstein coefficient
for spontaneous emission $A_{\rm ul}$ = 4.55$\times$10$^{-4}$ s$^{-1}$,
rotational constants as given in Tab.~4 of \cite{ah05}, lower state
energy of 50.2 K, and degeneracy of the upper and lower levels equal
to 9.

\begin{table}
\begin{center}
\caption{ p--\DTHP \ column density upper limits. \label{tdthp}}
\begin{tabular}{lccccccccc}
\hline\hline
Source &  $\Delta {\rm v}_{\rm res}$ & $T_{\rm sys}$ & $t_{\rm int}$ &
          $T_{\rm rms}$$^{\mathrm{a}}$ &
          $T_{\rm R}$ &  $\eta_{b}$ &  $T_{\rm ex}$$^{\mathrm{b}}$ & $N(p-\DTHP)$ &
          $p/o^{\mathrm{c}}$\\
Name   & (\kms ) & (K) & (min) & (K) & (K) & &  (K) & (10$^{13}$ cm$^{-2}$) & \\
\hline
B~1 & 0.083 & 3950 & 251 & 0.068 & $<$0.20 & 0.4  &  11.9 & $<$0.9 & $<$1.0 \\
NGC~1333--\DCOP & 0.083 & 3300 & 91 & 0.068 & $<$0.20 & 0.4  & 12.6 & $<$0.9 & $<$2.6 \\
NGC~2264G--VLA2 & 0.021 & 1436 & 93 & 0.045 & $<$0.14 & 0.4 & 14.4 & $<$0.2 & $<$0.7 \\
L~183 & 0.042 & 1770 & 42 & 0.084 & $<$0.25 & 0.4 & 6.9 & $<$7.9 & $<$3.2 \\
16293E$^{\mathrm{d}}$ & 0.10 & ... & 103 & 0.19 & 0.85 & 0.60 & 10.7 &  1.2\p 0.5 & 1.0 \\
L~1544$^{\mathrm{e}}$ & 0.021 & ... & 230 & 0.18 & $<$0.54 & 0.4 & 7.6 & $<$2.5 &
$<$0.8 \\
\hline
\end{tabular}
\begin{list}{}{}
\item[$^{\mathrm{a}}$] In radiation temperature units.
\item[$^{\mathrm{b}}$] Excitation temperature assuming a critical density of
10$^5$ \percc \ for the observed \DTHP \ transition.
\item[$^{\mathrm{c}}$] $p/o$ $\equiv$ $N(para-\DTHP )/N(ortho-\HTDP)$.
\item[$^{\mathrm{d}}$] Data from \cite{vpy04}. The column density has been
estimated using the method described in the text.
\item[$^{\mathrm{e}}$] Data from \cite{vcc06}.  The column density has been
estimated using the method described in the text, with the parameters
listed in Tab.~\ref{tcolumn}.
\end{list}
\end{center}
\end{table}

The last column of Tab.~\ref{tdthp} shows the ratio between the upper-limit
column densities of para-\DTHP \ and the ortho-\HTDP \
column densities listed in Tab.~\ref{tcolumn}.  The estimated values
are well within those calculated by \citet{fpw04} (see their figure
7) for cloud cores with volume densities 2$\times$10$^6$~\percc \ and
temperature ranges between 10 and 15~K.

\section{Analysis}
\label{sana}

\subsection{Derivation of the average ortho-\HTDP \  column densities}
\label{scolumn}

In this section we estimate the average ortho--\HTDP \ column
density in each source, assuming that the line is emitted in
homogeneous spheres at the density and temperature respectively quoted
in the literature and reported in Tab.~\ref{tcolumn}. Values of the \MOLH \ column densities are also reported in Tab.~\ref{tcolumn}, column 5, and they are used to determine the fractional abundance of ortho--\HTDP \ (see Sect.~\ref{s_correlations}). The $N(\MOLH )$ values for L~1544, L~429 and L~694-2 have been determined by \citet{ccw05} assuming constant temperature. However, a temperature gradient has been measured toward L~1544 (\citealt{ccw07}) and assumed (because of the similar physical structure) in L~429 and L~694-2, so that $N(\MOLH )$ needs to be modified. As found by \citet{pbm04} and \citet{sww07}, the temperature drop in L~183 and Oph~D implies an increase in $N(\MOLH )$ by a factor of about 1.4. Given that L~1544, L~429 and L~694-2 have structures similar to L~183 and Oph~D, we simply multiplied the \citet{ccw05} values by the same correction factor (1.4) to account for the temperature gradient and this is what is reported in Tab.~\ref{tcolumn}. 

We further assume that the level structure
of the ortho--\HTDP \ molecule is reduced to a two--level system. This 
is especially justified in starless cores, considering that the 
first excited state is 17.4~K above ground and the second one is 110~K. 
 In star forming regions, we assume that the observed 
\HTDP \ line is arising from gas with characteristics not significantly 
different from starless cores (which is probably true in L~1521F and 
NGC~2264G--VLA2, see Sect.~\ref{sres}). In the case of B~1 and NGC~1333--\DCOP ,
the broad lines suggest that the embedded young stellar objects have 
probably increased the degree of turbulence in the region and may have 
locally altered the conditions where \HTDP \ emits.  However, we should point
out that where the gas temperature increases above 20~K, and/or where 
shocks are present, \HTDP \ should not survive for long, considering 
that in these conditions the backward reaction (1) can quickly proceed and that dust mantles 
can be either evaporated or sputtered back into the gas phase, with the
consequence of increasing the CO abundance and thus the destruction rate
of \HTDP . 

To estimate the average ortho--\HTDP \ column density, we evaluate the
excitation temperature T$_{\rm ex}$ and the line optical depth $\tau$
simultaneously, by solving iteratively the following equations (valid for $T_{bg} \ll E_{ul}/k$):

\begin{eqnarray}
  T_{\rm ex} & = & \frac{T_{\rm
      kin}}{\frac{T_{\rm kin}}{E_{\rm ul}}
    ln \left( 1 + \frac{\beta n_{\rm cr}}{n_{\rm tot}} \right) +1}
\label{etex}
\end{eqnarray}
\begin{eqnarray}
\beta & = & \frac{0.75}{\tau}\left(1+\frac{e^{-2\tau}}{\tau}+ 
      \frac{(e^{-2\tau}-1)}{2\tau^2}\right)
\label{beta}
\end{eqnarray}
\begin{eqnarray}
\tau & = & - ln \left[ 1 - \frac{T_{\rm
        mb}}{J_{\nu}(T_{\rm ex}) - J_{\nu}(T_{\rm bg})} \right]
\label{tau}
\end{eqnarray}

\noindent
Eq.~(\ref{etex}) is the definition of T$_{\rm ex}$ corrected for the
line opacity: in practice, the critical density is reduced by the
probability that the emitted photon can indeed escape absorption.  We
adopted the photon escape probability $\beta$ of Eq. \ref{beta}, valid
for an homogeneous sphere \citep{ost89}. Finally the line optical
depth $\tau$ is derived from the observation via equation~(\ref{tau}).  
In the above equations, $E_{\rm ul}$ is the energy of the
transition ($E_{\rm ul}/k_{\rm B}$ = 17.4~K, with $k_{\rm B}$ the 
Boltzmann constant), 
$n_{\rm tot}$ is the particle volume density (assumed to be 1.2$\times
n(\MOLH )$, to account for He), T$_{\rm kin}$ is the gas temperature,
and $J_{\nu}(T_{\rm ex})$ and $J_{\nu}(T_{\rm bg})$ are the equivalent
Rayleigh--Jeans excitation and background temperatures. 

The ortho--\HTDP \ column density is then derived from $\tau$:
\begin{eqnarray}
N(ortho-H_2D^+) & = & \frac{8\pi\nu^3}{c^3}\frac{Q(T_{\rm ex})}{g_uA_{ul}}
        \frac{e^{E_u/T_{\rm ex}}}{e^{E_u/T_{\rm ex}}-1}
        \int \tau dv
\label{coldens}
\end{eqnarray}

\noindent
A key parameter is the critical density n$_{\rm cr}$ of the
transition. Recent and unpublished calculations by E.Hugo and
S.Schlemmer (private communication) suggest a collisional coefficient with ortho and para \MOLH \ 
of $\sim10^{-9}$ cm$^3$s$^{-1}$ at 10 K, with a very shallow dependence on 
the temperature. The implied critical density is
$\sim 10^5$ cm$^{-3}$, namely a factor of 10 lower than assumed in 
previous work \citep{bvw90,vcc05}. The ortho--\HTDP \ 
column densities, calculated both with the new and old critical density values,  
are reported in Tab.~\ref{tcolumn}.   The lines are either optically thin 
(e.g. in B~68) or marginally thick ($\simeq$1, in L~1544, L~183, L~694-2, L~429).
From the table, it is clear that a factor of 10 variation in the collisional coefficient
implies a change in the column density value of factors between 1.4 and 4.5, 
depending on the volume density and kinetic temperature. In the following 
analysis we use the column densities calculated with the 
10$^5$~cm$^{-3}$ critical density.  

The four objects which show the largest values 
of $N(ortho-\HTDP )$ are among the most centrally concentrated cores in the 
sample:  L~429, L~1544,  L~694-2 and L~183.  The two other dense cores, TMC--1C 
and Oph~D have significantly lower ortho--\HTDP \ column densities, and 
this may be related to different evolutionary stages.  We will further 
discuss these issues in Sect.~\ref{schem}.    

\subsection{Evaluation of the errors}

The estimates of the ortho--\HTDP \ column densities reported in 
Tab.~\ref{tcolumn} suffer from several sources of uncertainties. 
   We already mentioned a
basic source of uncertainty, that associated with the collisional
coefficient of the transition.  In addition to that, the densities and
temperatures used to derive the excitation temperatures are also
relatively uncertain, not only because of the uncertainty in deriving
these values at the centers of the studied sources but also because
the \HTDP \ line emission may originate in denser than the quoted
average density gas due to the \HTDP \ abundance distribution. In this
context, the errors associated with the rms of the observations
reported in Tab.~\ref{tfit} are certainly the smallest in the error
propagation chain. Although it is difficult to exactly quantify the
error in the determination of the ortho--\HTDP \ column densities, we
estimate here how reasonable changes in the gas temperature and density
would affect the reported column densities. Increasing or decreasing
the density by a factor of 2 results in decreasing/increasing the
ortho--\HTDP \ column densities by less than 30\%.  However, a change
in the kinetic temperatures of Tab.~\ref{tcolumn} by 1~K would change the
ortho--\HTDP \ column densities by up to a factor of 2 in the coldest
objects (because of the exponential in the level population equation).

In summary, considering also the beam efficiency variation between 0.4 and 
0.7 (Tab.\ref{tfit}), the ortho--\HTDP \ column densities reported in 
Tab.~\ref{tcolumn} are likely to be uncertain by about a factor of $\simeq$2.
 
\subsection{Correlations}
\label{s_correlations}

\begin{figure}
\includegraphics[angle=-90,scale=.50]{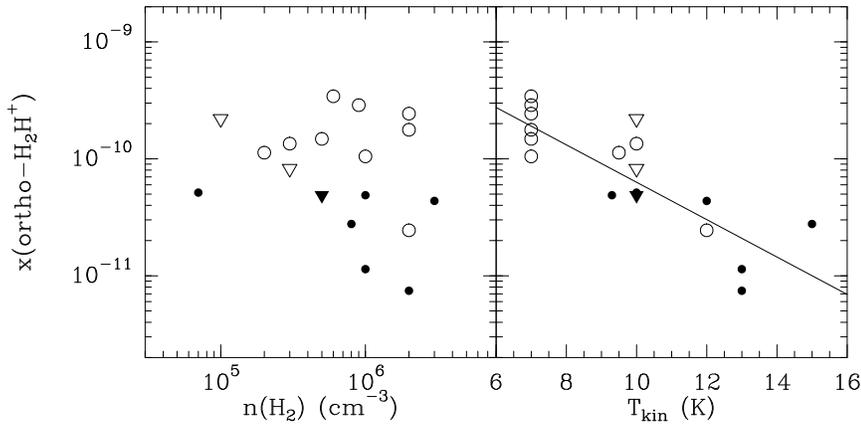}
\caption{ortho-\HTDP \ fractional abundances  vs. the gas volume density ($n(\MOLH )$, left panel) 
and gas temperature ($T_{\rm kin}$, right panel).  Empty symbols refer to starless
cores, whereas filled symbols refer to cores associated with young
stellar objects.  Note that upside--down triangles are upper limits.
The line in the  right panel is the least square fits to the $x$(ortho-\HTDP) vs.
T$_{\rm kin}$ data, the only significant correlation that has been found (see text).}
\label{fcorr}
\end{figure}

We have looked for possible correlations between the column density or 
fractional abundance of ortho-\HTDP  \ and  physical parameters such as 
the volume density, the \MOLH \ column density, the kinetic temperature
and the non--thermal line width.  No significant correlations have been found,
with the exception of $x(ortho-\HTDP )$ ($\equiv$ $N(ortho-\HTDP )/N(\MOLH )$, with 
$N(\MOLH )$ from Tab.~\ref{tcolumn}) vs. 
$T_{\rm kin}$, for which we find (see Fig.~\ref{fcorr}):

\begin{eqnarray}
Log x(ortho-\HTDP ) & = & (-8.6 \pm 0.3) - (0.16 \pm 0.03) T_{\rm kin},
\end{eqnarray}

\noindent
with a correlation coefficient of -0.83. 
What is causing the ortho-\HTDP  drop observed between 7~K and 15~K?   Partially, this
may be due to the fact that the warmest sources: ({\it i}) may have a smaller ortho-\HTDP \  
emitting region than found in  L~1544 (about 60$^{\prime\prime}$; \citealt{vcc06}), 
because of a drop of the ortho-\HTDP \ abundance (see Fig.~6 of \citealt{fpw04}) and ({\it ii}) 
are all at distances $>$ 300 ~pc (with the exception of 16293E, which, by the way, follows the trend). 
Thus, the \HTDP \ emission may be affected by beam dilution (not considered in this study). 
However, as we will see in the next section,  variations in the CO depletion factor, linked
to the gas density and, in turn, to the gas and dust temperature (at least in starless cores, 
larger temperatures are typically associated with lower gas densities, lower CO depletion
factors, and lower deuterium--fractionations),  can also (at least partially) produce the observed
trend.  In any case, a more detailed physical and chemical structure (such as that 
recently developed by \citealt{awg08}) is needed to investigate these
points, although the lack of information on the extent and morphology of the \HTDP \ emission  
prevents us from a more quantitative analysis of the correlations between the gas traced by
\HTDP \  and the physical properties of the selected cores. 

In the following we will concentrate on molecular abundances and use both a simple chemical model
applied to a homogeneous cloud and a slightly more detailed chemical-physical model of a centrally
concentrated and spherically symmetric cloud to reproduce and 
interpret the observed variations of ortho-\HTDP  column densities, deuterium fractionation 
and CO depletion fraction in the selected cores. 

\section{Chemical Discussion}
\label{schem}

The chemistry of starless cores and their degree of deuteration has been 
investigated in detail by \citet{rhm03}, \citet{fpw04}, \citet{rhm04}, \citet{wfp04}, \citet{ahr05}, 
\citet{fpw05}, \citet{fpw06a}, \citet{fpw06b}. From these models 
it appears that there are several sources of uncertainty that can profoundly 
affect the chemistry: (i) surface chemistry (diffusion and reaction rates 
are still quite uncertain and highly dependent upon the poorly known 
surface of dust grains); (ii) freeze--out and desorption rates (binding 
energies may be changing throughout the core, due to changes in grain mantle
composition,  and nonthermal desorption processes are poorly known; see 
\citealt{gpc06,gwh07}); (iii) the dynamical evolution of dense cores is hard to 
constrain chemically, and different theoretical models of star formation 
predict significantly different time scales (e.g. \citealt{sal87} and 
\citealt{hbb01}); (iv) the cosmic--ray ionization rate $\zeta$ is not 
well constrained, and our ignorance in the cosmic--ray energy spectrum 
(especially at low energies) prevents us from making quantitative estimates
on the possible variations of $\zeta$ within dense cores (see \citealt{ps05}, 
for a recent discussion on this point, and \citealt{d06} for a more general 
review); (v) the fraction of 
ortho--\MOLH,
which affects the deuteration in the gas phase 
(the backward reaction (1) proceeds faster with ortho--\MOLH , because of its
higher energy compared to para--\MOLH ; \citealt{ghr02}, hereafter GHR02; \citealt{wfp04}); 
(vi) the dust grain size distribution (if dust grains coagulate in the
densest regions of starless cores, the freeze--out rate diminishes, altering 
the gas phase chemical composition; see, e.g. \citealt{fpw06a,vcc06}); 
(vii) the abundance of polycyclic aromatic hydrocarbons
(PAHs), unknown in dark clouds (where observational constraints are yet to 
be found), which may significantly affect the 
electron fraction \citep{ld88, fp03,t05}; (viii) the fraction 
of atomic oxygen in the gas 
phase, which is thought to be low (mainly to explain the stringent SWAS
upper limits on the water abundance; e.g. \citealt{bs02}), but
observations of dark clouds with the ISO satellite appear to disprove this 
\citep{ccc99,vcc00,lkp01}, although the limited velocity 
resolution of ISO LWS is a severe limit on these results. In the following section, the effects of some of the above parameters on the deuterium fractionation will be presented for the simple case of a homogeneous cloud.  In Sec.~\ref{smod}, a more detailed physical structure and a slightly more comprehensive chemical model will be considered to make an attempt on constraining some of the unknown parameters.  

\subsection{Simple Theory: what affects deuterium fractionation}
\label{s_simple}

Ignoring for the moment the density and temperature
structure of molecular cloud cores and any gas-dust interaction, which will be considered in 
Sect.~\ref{smod}, we show here simple relations between the \HTDP /\HTHP \ 
abundance ratio and parameters such as the gas kinetic temperature, the 
grain size distribution, the CO depletion factor and the cosmic-ray 
ionization rate. 
%We use a simple chemical network which includes all the multiply 
%deuterated forms of \HTHP , which are formed following the reaction scheme 
%listed in Table~1 of \citet{cd05} (hereafter CD05), based on
%Roberts et al. (2004), and are destroyed by CO, electrons (dissociative
%recombination) and negatively charged dust grains (recombination). 
%The adopted rate coefficients are the same as in Table~1 of CD05, with the 
%exception of the reaction of \HTHP \ (and deuterated isotopologues) with  CO, and 
%the reaction of \HTHP \ and electrons, for which we used the values listed in the 
%UMIST database (http://www.udfa.net/). 
We use a simple chemical network which includes all the multiply deuterated forms of \HTHP , 
formed following the reaction scheme listed in Tab.~\ref{rate} and destroyed by CO, electrons 
(dissociative recombination) and negatively charged dust grains (recombination). The adopted rate 
coefficients are the same as in Tab.~1 of \citet{cd05}, with the exception of the reaction of \HTHP \ (and deuterated 
isotopologues) with  CO, and the reaction of \HTHP \ and electrons, for which we used the values listed in the 
latest release of the UMIST database (RATE06) available at http://www.udfa.net/.  We note that the rate coefficient of the 
\HTHP \ + CO reaction in the UMIST database is temperature-independent, 
as expected for ion--molecule reactions where the neutral species has a small
dipole moment (E. Herbst, private communication). 

\begin{table}
\begin{center}
\caption{The forward rate coefficient is given by $\alpha(T/300)^{\beta}$. The reverse rate is given by $\alpha(T/300)^{\beta}e^{-\gamma/T}$.\label{rate}}
\begin{tabular}{cccccc}
\hline
Reaction  &  Rate    & $\alpha$                 &  $\beta$    & $\gamma$ & Ref.\\
                  &               & cm$^3$~s$^{-1}$   &                    &       K           &              \\
\hline
H$_3^+$ + HD $\leftrightarrow$ H$_2$D$^+$ +H$_2$           & k$_1$, k$_{-1}$ & 1.7 10$^{-9}$ & 0 & 220 & (1)\\
H$_2$D$^+$ + HD $\leftrightarrow$ D$_2$H$^+$ +H$_2$   & k$_2$, k$_{-2}$ & 8.1 10$^{-10}$ & 0 & 187 & Roberts et al. (2004)\\
D$_2$H$^+$ + HD $\leftrightarrow$ D$_3^+$ +H$_2$           & k$_3$, k$_{-3}$ & 6.4 10$^{-10}$ & 0 & 234 & Roberts et al. (2004)\\
H$_3^+$ + CO $\rightarrow$ HCO$^+$ + H$_2$ & k$_{CO}$  & 1.7 10$^{-9}$ & 0 & 0 & RATE06\\
H$_2$D$^+$ + CO $\rightarrow$ HCO$^+$ + HD  & k$_{CO}$  &1.7 10$^{-9}$ & 0 & 0 & RATE06\\
\hspace{1.4cm} $\rightarrow$ DCO$^+$ + H$_2$  &   & &  &  & \\
D$_2$H$^+$ + CO $\rightarrow$ HCO$^+$ + D$_2$  & k$_{CO}$  &1.7 10$^{-9}$ & 0 & 0 & RATE06\\
\hspace{1.6cm} $\rightarrow$ DCO$^+$ + HD  &   & &  &  & \\
D$_3^+$ + CO $\rightarrow$ DCO$^+$ + D$_2$ & k$_{CO}$  &1.7 10$^{-9}$ & 0 & 0 & RATE06\\
H$_3^+$ + e$^-$ $\rightarrow$ $products$ & k$_{rec0}$  & 6.7 10$^{-8}$  & -0.52  &  0 & RATE06 \\
H$_2$D$^+$ + e$^-$ $\rightarrow$ $products$   & k$_{rec1}$  & 6.0 10$^{-8}$  & -0.5  &  0 & \citet{smd94} \\
D$_2$H$^+$ + e$^-$ $\rightarrow$ $products$  & k$_{rec2}$  & 6.0 10$^{-8}$  & -0.5  &  0 & \citet{rhm04} \\
D$_3^+$ + e$^-$   $\rightarrow$ $products$  & k$_{rec3}$  & 2.7 10$^{-8}$  & -0.5  &  0 & \citet{ldl97} \\
\hline
\end{tabular}
\end{center}
%\begin{description}
{(1) At the low temperatures found in our starless and protostellar cores, this rate corresponds to the Langevin limit.}
%\end{description}
\end{table}

The steady state equations for this simple system are:

\begin{eqnarray}
\frac{x(\HTDP )}{x(\HTHP )} & = & \frac{k_1 x({\rm HD}) [D_2 D_3 - x({\rm HD}) k_{-3} k_3]}
{D_1 D_2 D_3 - x({\rm HD}) (k_{-3} k_3 D_1 - k_{-2} k_2 D_3)} \label{e_dfrac1} \\
\frac{x(\DTHP )}{x(\HTDP )} & = & \frac{k_2 x({\rm HD}) D_3}{D_2 D_3 - 
k_{-3} k_3 x({\rm HD})} \label{e_dfrac2} \\
\frac{x(\DTHREEP )}{x(\DTHP )} & = & \frac{k_3 x({\rm HD})}{k_{-3} + k_{co} x({\rm CO})
+ k_{rec3} x(e) + k_g x(g)} \label{e_dfrac3} ,  
\end{eqnarray}

\noindent
with

\begin{eqnarray*}
D_1 & = & k_{-1} + k_{co} x({\rm CO}) + k_{rec1} x(e) + k_g x(g) + k_2 x({\rm HD}) \\
D_2 & = & k_{-2} + k_{co} x({\rm CO}) + k_{rec2} x(e) + k_g x(g) + k_3 x({\rm HD}) \\
D_3 & = & k_{-3} + k_{co} x({\rm CO}) + k_{rec3} x(e) + k_g x(g) .
\end{eqnarray*}
 
\noindent
In the above expressions, $k_1$, $k_2$, $k_3$ and $k_{-1}$, $k_{-2}$, $k_{-3}$ are 
the forward and backward rate coefficients relative to reactions of \HTHP , \HTDP , and \DTHP , 
respectively, with HD; $x(i)$ is the fractional abundance (w.r.t. \MOLH ) 
of species $i$; $x({\rm HD})$ = 3$\times$10$^{-5}$ is the 
fractional abundance of HD (assuming that [D]/[H] = 1.5$\times$10$^{-5}$, 
\citealt{ohh03}, and that in molecular clouds the deuterium is 
mainly in the form of HD); $k_{\rm CO}$  is the rate coefficient of the
destruction reaction of all \HTHP \ isotopologues with CO; 
$x({\rm CO})$ = $x_{\rm can}(\rm CO)/f_{\rm D}$, where $x_{\rm can}(\rm CO)$ = 
9.5$\times$10$^{-5}$ is the canonical abundance of CO as measured by 
\citet{flw82}, and $f_{\rm D}$ is the CO 
depletion factor (1/$f_{\rm D}$ is the fraction of CO molecules 
left into the gas phase, see Sect.~\ref{s_fd}); $k_{rec1}$, $k_{rec2}$, 
$k_{rec3}$ are the dissociative recombination rate coefficients for 
\HTDP , \DTHP , and \DTHREEP , respectively. Following 
\citet{ds87} (see also \citealt{ccw04}), the rate coefficient for the 
recombination onto dust grains has the form:

\begin{eqnarray}
k_{\rm g} & = & 1.6\times 10^{-7} \left( \frac{a_{\rm min}}{10^{-8} 
{\rm cm}} \right) \left( \frac{T_{\rm kin}}{10 {\rm K}} \right)^{-0.5} 
\times \left( 1 + 3.6\times10^{-4} \frac{T_{\rm kin}}{10 {\rm K}} 
\frac{a_{\rm min}}{10^{-8} {\rm cm}} \right) 
\end{eqnarray}

\noindent
where $a_{\rm min}$ (=50~\AA) is the minimum radius of dust grains in the 
\cite{mrn77} (MRN) size distribution ($a_{\rm max}$ = 
2.5$\times$10$^{-5}$ cm).  Finally, the fractional 
abundance ($x( g)$) of dust grains in a MRN size distribution is given by:

\begin{eqnarray}
x_{\rm g} & = & 5.3 \times 10^{-6} \left( \frac{a_{\rm max}}{10^{-4} {\rm 
cm}} \right)^{-0.5} \left( \frac{a_{\rm min}}{10^{-8} {\rm cm}} 
\right)^{-2.5}.
\end{eqnarray}

\noindent
The electron fraction is calculated as in \citet{cwz02b}, using 
a simplified version of the reaction scheme of \citet{un90}, 
where we compute a generic abundance of molecular ions ``mH$^+$'' assuming
formation due to proton transfer with \HTHP \ and destruction by 
dissociative recombination on grain surfaces (using rates from \citealt{ds87}). 
\HTHP \ is formed as a consequence of cosmic--ray ionization
of \MOLH \ and destroyed by proton transfer with CO and \MOLN . Metals 
are also taken into account and their fractional abundance has been 
assumed 10$^{-7}$ (following the initial abundances of "low-metal" models, appropriate
for dark clouds; \citealt{lbh96}). 

The deuterium fractionation in species such as \HCOP \ or \NTHP \ 
($R_{\rm D}$ $\equiv$ [\DCOP ]/[\HCOP ] or [\NTDP ]/[\NTHP ]) in this 
chemical scheme is simply given by:

\begin{eqnarray}  
R_{\rm D} & = & \frac{1/3 x(\HTDP ) + 2/3 x(\DTHP ) + 
              x(\DTHREEP)}{x(\HTHP ) + 2/3 x(\HTDP ) + 1/3 x(\DTHP )} .
\label{erd}
\end{eqnarray}

\begin{figure}
\includegraphics[angle=-90,width=1.0\textwidth]{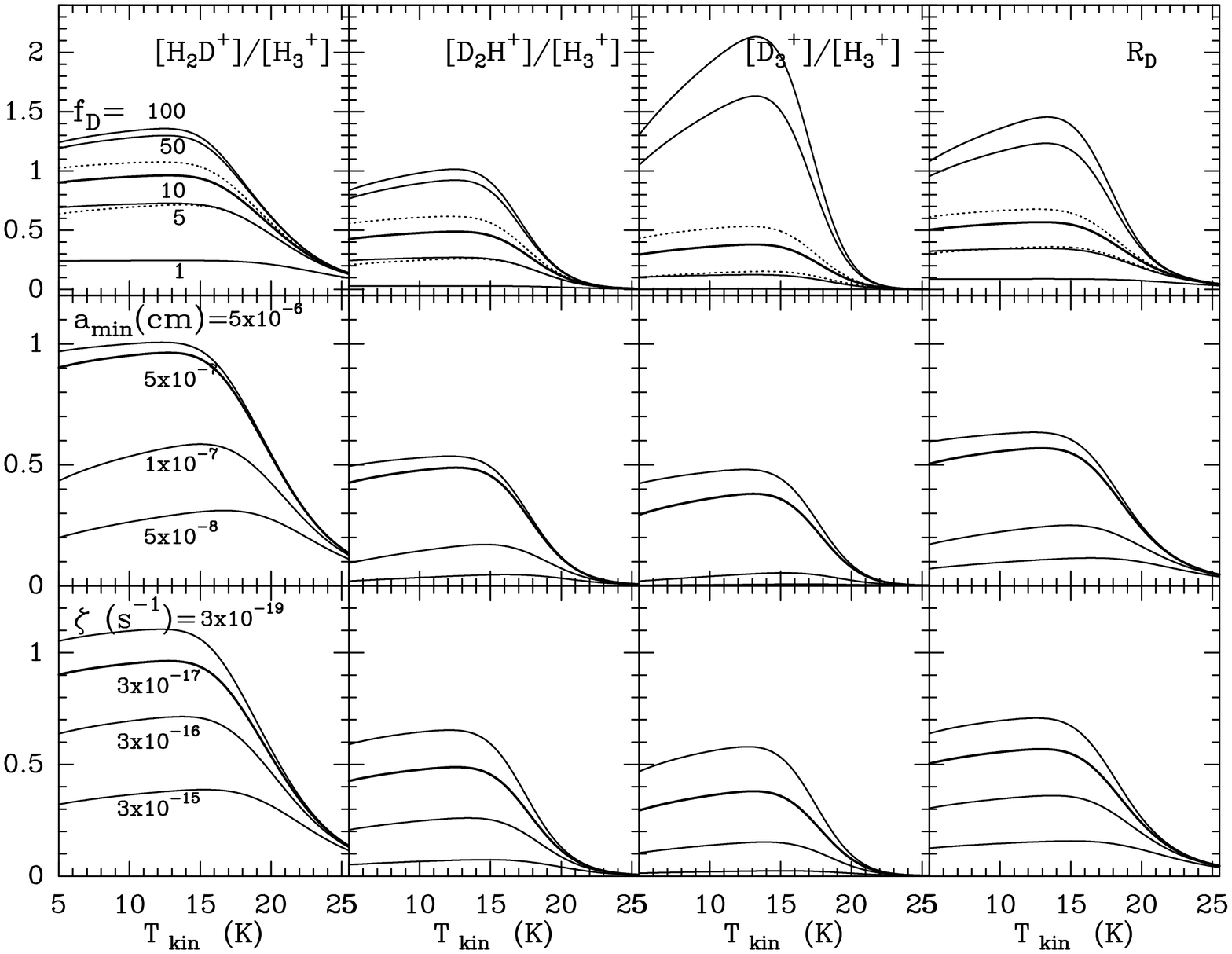}
\caption{[\HTDP ]/[\HTHP ], [\DTHP ]/[\HTHP ], [\DTHREEP ]/[\HTHP ] and 
$R_{\rm D}$ abundance ratios as a function of the gas temperature
($T_{\rm kin}$) in a dense cloud with uniform volume density $n(\MOLH )$ =
10$^5$~\percc .  ({\it Top row}) The abundance ratios are plotted against 
$T_{\rm kin}$ for different values of the depletion factor $f_{\rm D}$ (=1, 5,
10, 50 and 100), with fixed values of $a_{\rm min}$ = 50~\AA~and 
$\zeta$ = 3$\times$10$^{-17}$ s$^{-1}$. The dotted curves are for $f_{\rm D}$ = 10 clouds with
$n(\MOLH )$ = 1$\times$10$^6$ \percc \ (top dotted curve) and $n(\MOLH )$ =
1$\times$10$^4$ \percc \ (bottom dotted curve). Note the large increase 
of the [\DTHREEP ]/[\HTHP ] ratio for $f_{\rm D}$ = 50 and 100. 
({\it Central panel}) Abundance ratios vs. $T_{\rm kin}$ for different values of
$a_{\rm min}$ and values of $f_{\rm D}$ and $\zeta$ fixed at 10 and 
3$\times$10$^{-17}$~s$^{-1}$, respectively. ({\it Bottom panel})
Abundance ratios vs. $T_{\rm kin}$ for different values of the cosmic--ray
ionization rate, $\zeta$. Here, $f_{\rm D}$ = 10 and $a_{\rm min}$  =
50~\AA.}
\label{fdeuteration}
\end{figure}

Fig.~\ref{fdeuteration} shows the [\HTDP ]/[\HTHP ], [\DTHP ]/[\HTHP ], 
[\DTHREEP ]/[\HTHP ] and $R_{\rm D}$ abundance ratios
as a function of gas temperature, for different 
values of the depletion factor $f_{\rm D}$ (top panel), the minimum 
value of the dust grain radius in the grain--size distribution, $a_{\rm min}$
(central panel), and the cosmic--ray ionization rate, $\zeta$ (bottom panel).  
The first thing to note is the sharp drop in the deuterium
fractionation at temperatures above $\simeq$15~K, when reaction 
(\ref{e_htdp}) and the analogous ones for the formation of \DTHP \ and 
\DTHREEP \ also start to proceed backward.  Between 5 and 10~K, the 
deuteration ratios increase because of the inverse temperature 
dependence of the \HTDP \ destruction rate coefficients 
$k_{\rm rec1}$, $k_{\rm rec2}$, $k_{\rm rec3}$ and $k_g$ (present in the denominator 
of eqs.~\ref{e_dfrac1}, 9, and 10).  Note the high peaks in the $f_{\rm D}$ = 50 and 100 
[\DTHREEP ]/[\HTHP ] curves.  This is due to the fact that in regions 
where most of the neutrals are frozen (although here we only talk about CO, 
we can generalize this statement including in the CO--depletion factor 
all the neutral species which react with the \HTHP \ isotopologues), 
the deuterium fractionation proceeds rapidly and \HTHP \ is efficiently converted into \DTHREEP , as 
already shown by \citet{wfp04} in the case of dense cloud cores and by \citet{cd05} 
in the case of protoplanetary disks. The large abundances of \DTHREEP \ 
yield large $R_{\rm D}$ values ($\geq$ 1).

Fig.~\ref{fdeuteration} also shows that at temperatures 
below $\simeq$17~K, the deuterium fractionation is very much dependent 
upon the CO depletion factor (a well known phenomenon; e.g. \citealt{dl84}), 
the gas volume density (see the dotted curves in the top figures), 
the fractional abundance of dust grains and the cosmic--ray ionization rate.  
In particular, a value of $a_{\rm min}$ = 5$\times$10$^{-8}$ cm (5~\AA ), 
corresponds to $x(g)$ = 2$\times$10$^{-7}$, which may be regarded as 
a possible value for the fractional abundance of PAHs 
\citep[e.g.][]{ld88,t05}. 
Thus, if PAHs are abundant in dense cores and they are the main 
negative charge carriers, the deuterium fractionation is expected 
to be $\leq$0.1, because the four \HTHP \ isotopologues quickly recombine. 
The fact that molecules such as \NTHP \ and \AMM \ show large deuterium fractionations
(or large $R_{\rm D}$ values) in the direction 
of pre-stellar cores and Class 0 sources ($R_{\rm D}$ $>$ 0.1; see 
Fig.~\ref{fchem01}),  thus suggests that (negatively charged) PAHs have 
abundances significantly below 10$^{-7}$.  

The bottom panel of Fig.~\ref{fdeuteration} shows that the larger 
the cosmic--ray ionization rate the smaller the deuteration ratios. 
This is mainly due to the fact that a larger value of $\zeta$ implies 
a larger electron fraction, and a consequently larger dissociative recombination
rate (see denominator of eqs.~\ref{e_dfrac1}, 9, and 10).   Again, the large deuterium 
fractionation observed in pre--stellar cores and Class 0 objects can be used 
to put upper limits on $\zeta$ \citep[see][]{d06}. 

Before proceeding to the next sub--section, we note here that the $R_{\rm D}$ 
values obtained in this analysis for typical parameters ((i) $f_{\rm D}$ $\simeq$ 10, 
as typically observed in pre--stellar cores, (ii) $a_{\rm min}$ = 50~\AA , as in 
the MRN distribution, and (iii) $\zeta$ $\simeq$ 3$\times$10$^{-17}$~s$^{-1}$) reach 
$\simeq$0.5 for $T_{\rm kin}$ $\leq$ 15~K. This value may appear too large when compared to 
the deuterium--fractionation measured in pre--stellar cores by \citet{ccw05}, but it is quite close to the 
[NH$_2$D]/[\AMM ] ratio found by (1) \citet{ccw07} in the nucleus of the pre--stellar core L~1544
using interferometric observations and by (2) \citet{pwh07} in Infrared Dark Clouds. 
However, the results presented in this section apply to an ideal homogeneous cloud 
and are based on ``standard'' rate coefficients for the proton--deuteron 
exchange reactions \HTHP , \HTDP , \DTHP +  HD. In fact, GHR02 have 
measured slower rates which, if adopted, lead to 
$R_{\rm D}$ values about a factor of 3 lower compared to those obtained in 
Fig.~\ref{fdeuteration}. This will be discussed in the next sub--section.

\subsection{Simple Chemical--Physical Model}
\label{smod}

In this section,  our estimates of ortho--\HTDP \ 
column densities are correlated
with the deuterium fractionation and the depletion factor, 
previously measured in the same objects, and simple chemical models are used 
to investigate the observed variations among the various sources. 
The model used is similar to that described in \citet{vcc06}, 
and first applied by \citet{cwz02b} in the case of L~1544, but with the 
deuterium fractionation chemistry and rate coefficients 
as described in the previous sub--section. 
We consider a spherical cloud, with a given density and temperature 
profile, where dust and gas are present. Initially (besides \MOLH ), 
CO and N$_2$ are 
present in the gas phase with abundances of 9.5$\times$10$^{-5}$ 
\citep{flw82} and 3.75$\times$10$^{-5}$, respectively. 
The abundance of \MOLN \ assumes that 50\% of the nitrogen is in 
atomic form (but no Nitrogen chemistry is considered, except for the \MOLN \ adsorption/desorption onto/from dust grains and the formation and destruction of \NTHP \ and \NTDP ). This is a totally arbitrary choice, but the \MOLN \ abundance is
extremely uncertain (see e.g. \citealt{sab03,fpw06b}; \citealt{mbl06}) and in any case, 
its variation in the gas phase only affects, in our simple model, the
absolute abundance of \NTHP , without significantly affecting the \NTDP /\NTHP 
column density ratio.  Atomic oxygen has not been included in the 
chemistry, in order to avoid one extra (uncertain) parameter in the model. It 
is worth to point out here that adding atomic oxygen to the chemical network lowers 
the deuterium fractionation, if its binding energy (also not well 
constrained) is sufficiently low (see e.g. \citealt{cwz02b} and their
discussion in Sect. 3.2).

The dust grain distribution follows MRN and we 
assume a gas--to--dust mass ratio of 100. However, the 
size of the minimum dust radius, $a_{\rm min}$, has been increased 
by an order of magnitude, following recent (indirect evidences) of grain growth 
toward the center of dense cores \citep[e.g.][]{fpw05,bmv06,fpw06b,vcc06}.  The higher $a_{\rm min}$ adopted here 
(5$\times$10$^{-6}$ cm) lowers the freeze--out rate by a factor of 5 
compared to the MRN value (by changing the surface area of dust grains), 
and it is the ``best--fit'' value for the L~1544 chemical model 
\citep[see][]{cwz02b,vcc06}. The freeze--out time scale of species $i$ ($R_{\rm freeze}(i)$) is given by:
\begin{eqnarray}
t_{\rm freeze}(i) & = & \frac{1}{S \, \Sigma \, <v_{\rm th}(i)> n_{\rm H}} 
    \nonumber \\
& = & \frac{2 \times 10^{4} {\rm yr}}{S}  \times  
\left[ \left( \frac{a_{\rm min}}{10^{-5} {\rm cm}} \right)^{-0.5} - 
\left( \frac{a_{\rm max}}{10^{-5} {\rm cm}} \right)^{-0.5} \right]^{-1} 
\times \nonumber \\
& & A(i)^{0.5} \left( \frac{10 {\rm K}}{T_{\rm gas}} \right)^{0.5} 
\left( \frac{10^5 {\rm cm^{-3}}}{n_{\rm H}} \right)
\label{efre}
\end{eqnarray}
where $S$ is the sticking coefficient ($\simeq$ 1, as recently found 
by \citealt{bfo06}), $<v_{\rm th}(i)>$ is the average thermal velocity of 
species $i$, $n_{\rm H}$ is the total number density 
of H nuclei ($n_{\rm H}$ = $n$(H) + 2$n(\MOLH )$), $A(i)$ is the 
atomic mass number of species $i$, and 
\begin{eqnarray}
\Sigma & = & (4.88\times 10^{-25} + 4.66\times 10^{-25}) \times 
 (a_{\rm min}^{-0.5} - a_{\rm max}^{-0.5}) ,
\end{eqnarray}
is the grain surface area per H nucleon in the MRN distribution (see 
also \citealt{wd01}).  The electron fraction is calculated as in Sect.~\ref{s_simple}.

The binding energies for CO and \MOLN \ have been taken from \citet{ovf05},
assuming that the mantle composition is a mixture of CO and H$_2$O 
ice ($E_{\rm D}$(CO)/$k_{\rm B}$ 
= 1100~K and $E_{\rm D}$(\MOLN ) = 0.9 $E_{\rm D}$(CO)). The cosmic--ray
ionization rate used here is $\zeta$ = 1.3$\times$10$^{-17}$ s$^{-1}$, 
but we have also changed it to explore the effects on the chemistry (see 
next subsections).  Different density structures  have been considered 
(see below) and the (gas = dust\footnotemark{} 
\footnotetext{This assumption of similar 
dust and gas temperatures is only valid if the densities are larger than 
$\sim$10$^5$ \percc \  \citep[e.g.][]{g01}. However, in the range of 
temperatures typical of low--mass dense cores, this approximation 
does not significantly change the results of our model.}) 
temperature profile is similar to the one found by \citet{yle04}
and parametrized so that:
\begin{eqnarray}
T_{\rm dust} \sim T_{\rm gas} & \simeq & 3 \times [8.7 - log(n(\MOLH )] 
\,\, {\rm K} .
\label{etem}
\end{eqnarray}
The minimum (maximum) allowed temperature is 4~K (14~K). We also 
consider models with temperature profiles increasing inwards, to simulate 
the heating of the embedded young stellar object.
 In this case, we assume a central temperature 
of 50~K at a distance of 80~AU, and a radial dependence $r^{-0.6}$ for 
$r > 80$~AU.  If the temperature drops below the one described 
in eq.(\ref{etem}), the latter value is used.  

CO and \MOLN \ can freeze--out (with rates given by 1/$t_{\rm freeze}$, 
see eq.~\ref{efre}) 
and return to the gas phase via thermal desorption or cosmic--ray impulsive
heating (following \citealt{hhl92} and \citealt{hh93}). The 
abundance of the molecular ions (\NTHP , \HCOP , \HTHP \ and their deuterated
isotopologues) are calculated in terms of the instantaneous abundances of 
neutral species (assumption based on the short time scale of ion--chemistry, 
compared to the depletion time scale; see \citealt{cwz02b} for details). 

\subsubsection{$N(ortho-\HTDP )$ vs. the observed $R_{\rm D}$}
\label{ss_htdp_rd}

\begin{figure}
\centering
\includegraphics[width=0.4\textwidth]{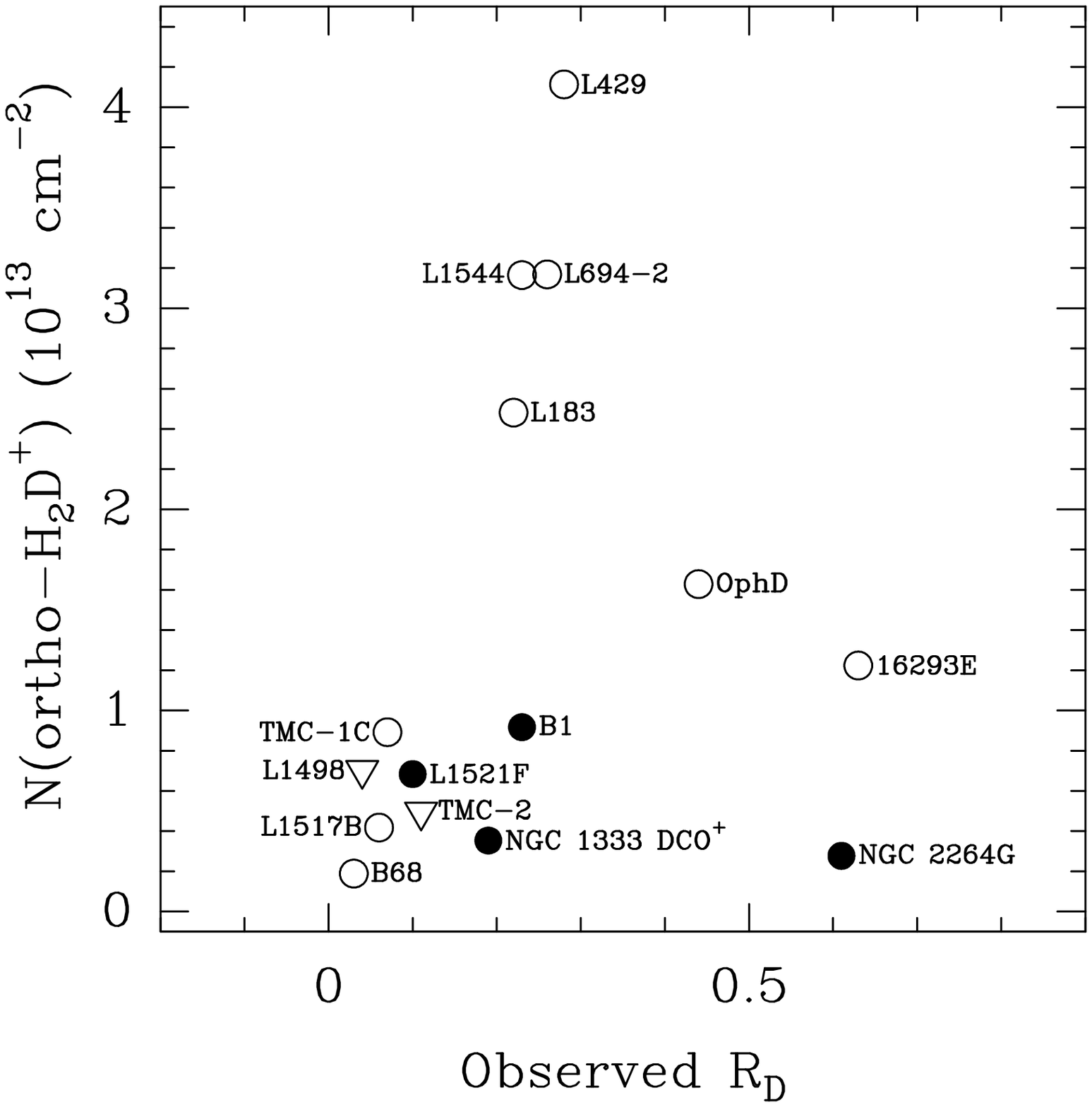}
\quad
\includegraphics[width=0.425\textwidth]{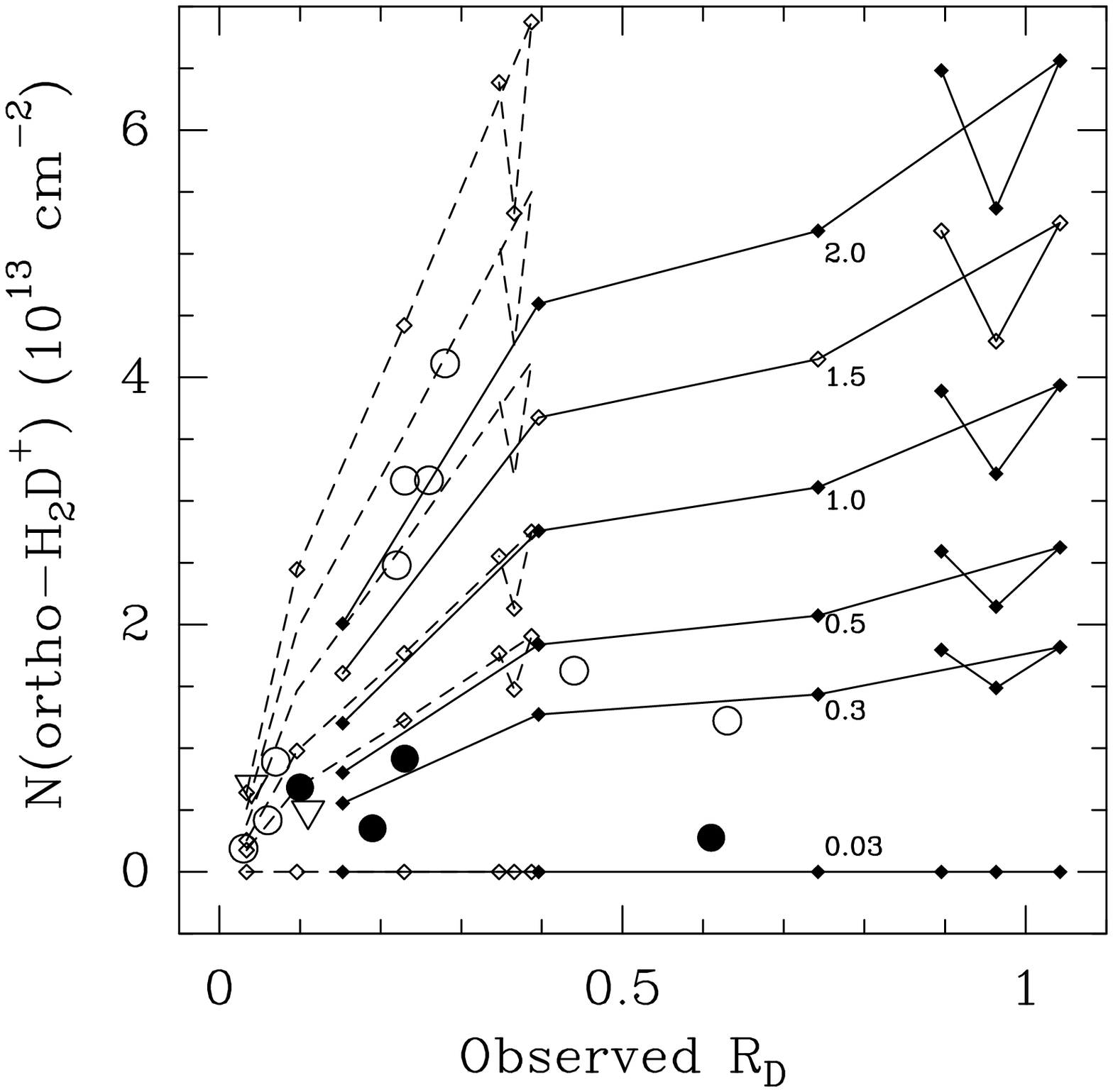}
\caption{ortho--\HTDP \ column density as a function of the observed
deuterium--fractionation $R_{\rm D}$ (see text for details). Filled circles are
protostellar cores, whereas
empty circles are starless cores. Downward arrows denote upper limits
in our estimate of the column density.  The names of the cores associated
with each mark in the figure are shown in the left panel.  The right panel
shows the same plot, where theoretical predictions from chemical models
(see text) are superposed.  Solid curves are for models which use the 
``standard'' rate coefficients for the proton--deuteron exchange reactions, 
whereas dashed curves are for models adopting the smaller GHR02 rate coefficients.
Each curve has six points (filled (empty) diamonds for the models using the "standard" (GHR02) rate coefficients), which correspond
to model cores with different central density (from 4$\times$10$^4$~cm$^{-3}$
to 4$\times$10$^8$~cm$^{-3}$, from left to right).  Different curves are for different
ortho/para \HTDP \ ratios, from 0.03 to 2.0. The cosmic--ray ionization rate has been fixed
to 1.3$\times$10$^{-17}$~s$^{-1}$.}
\label{fchem01}
\end{figure}

In Fig.~\ref{fchem01}, the column density of ortho--\HTDP \ is 
plotted as a function of the observed deuterium fractionation ratio ($R_{\rm D}$).
$R_{\rm D}$ is equivalent to  $N(\NTDP ) /N(\NTHP )$, in the case of starless 
cores (plus L~1521F), 
and this value has been taken from the survey of \cite{ccw05}.
In star forming regions, the  \NTDP /\NTHP \ column density 
ratio is not available, and other column density ratios have been used:
(i) $N({\rm NH_2D})/N({\rm NH_3})$ for NGC~1333--\DCOP \ 
\citep{h03} and for B~1 \citep{rlv05}; 
(ii) $\sqrt{N({\rm D_2CO})/N({\rm H_2CO})}$ for NGC~2264G~VLA2 \citep{lcc02}. 
No deuterium fractionation estimates are available for IRAM~04191.
Given that \AMM \ and \NTHP \ appear to trace similar zones of 
dense cores \citep[e.g.][]{bcm98,cbm02}, and that 
both derive from the same parent species (\MOLN ), one expects that the 
D-fractionation observed in the two species is also similar (and 
linked to the theoretical $R_{\rm D}$ in eq.~\ref{erd}). However, it is not
obvious that formaldehyde is actually tracing the same region (indeed 
H$_2$CO is centrally depleted in the two starless cores studied by 
Tafalla et al. 2006, unlike \NTHP \ and \NTDP ). In fact, Fig.~\ref{fchem01} shows that the 
deuterium--fractionation in NGC~2264G is the largest one in the whole sample, 
probably suggesting that different deuteration mechanisms (beside the \HTHP \ 
fractionation) may be at work for H$_2$CO. One possibility is that surface chemistry 
is needed to explain the observed amount of
deuterated formaldehyde and methanol, as originally discussed 
by \citet{ctr97}, \citet{ccl98}, and more recently by \citet{pct06}.

In the left panel, each data point is labelled with the corresponding 
name, whereas the right panel shows the same data points with 
model curves superposed (see below). The first thing to note is that, on average, 
dense protostellar cores (filled symbols) have lower $N(ortho-\HTDP )$ 
values than starless cores, 
but on average they show quite large deuterium fractionations (especially
in the case of NGC~2264G~VLA2, where $R_{\rm D}$ is coming from measurements
of doubly deuterated formaldehyde, as already mentioned).  Another thing to 
note is that there is {\it not} any clear correlation between 
$N(ortho-\HTDP )$ and the observed $R_{\rm D}$.  To investigate this 
unexpected result, we used the model described in Sec.~\ref{smod} and 
simulate an evolutionary sequence, similar to what has been done 
in \citet{ccw05} in their Fig.~5.  

We consider Bonnor-Ebert (BE) spheres with density structures analogous to the 
radial (cylindrical) density profile of the contracting disk--like cloud 
at different stages of evolution in the model of 
\cite{cb00}, namely those at times $t$ = $t_1$ (=2.27 Myr
and central densities $n_{c1}(\MOLH )$ = 4$\times$10$^4$ \percc ), 
$t_2$ (=2.60 Myr, and  $n_{c2}(\MOLH )$ = 4$\times$10$^5$ \percc ), 
$t_3$ (=2.66 Myr, and $n_{c3}(\MOLH )$ = 4$\times$10$^6$ \percc ),
$t_4$ (=2.68 Myr, and $n_{c4}(\MOLH )$ = 4$\times$10$^7$ \percc ), and 
$t_5$ (=2.684 Myr, and $n_{c5}(\MOLH )$ = 4$\times$10$^8$ \percc ). The 
BE density profile is reasonably well reproduced by the 
parametric formula \citep{tmc02}:
\begin{eqnarray}
n(r) & = & \frac{n_c(\MOLH )}{(1 + (r/r_0)^{\alpha})}
\label{eden}
\end{eqnarray}
where $r_0$ = 13,000, 3,000, 800, 300, and 80 AU for $t_1$, $t_2$, 
$t_3$, $t_4$, and $t_5$, respectively, and $\alpha$ = 2. The five chemical
models (i.e. the CO and N$_2$ depletion chemistry in the five model clouds with different physical structure) have been run for a time interval given by ($t_i$ - $t_1$), for 
$i$ = 2, 3, 4, and 5, whereas the $t_1$ model has been run for 2.27 Myr. 
We also consider a model cloud with the same density 
structure as the $t_5$ model, but with a temperature profile 
resembling that of a centrally heated protostellar core (model $t_{5a}$), with a central temperature of 50~K and a temperature gradient proportional to $r^{-0.6}$ (as mentioned above). 
The abundance profiles of \HTDP , \NTHP \ and 
\NTDP \ calculated by the models
have been convolved with 22\arcsec , 26\arcsec , and 17\arcsec \
FWHM antenna beams, respectively, to simulate the observations and 
calculate the column densities toward the core center (as well as off peak).  
 
The results of these models are the small diamonds in each of the 
curves in the right panel of Fig.~\ref{fchem01}, with $t_1$ lying on the 
left--end and $t_{\rm 5a}$ on the right--end of the curve.  
Solid curves represent
models with standard rate coefficients for the proton--deuteron 
exchange reactions, whereas dashed curves models use the 
about 3 times smaller GHR02 rates (see previous sub--section).     
The different curves correspond to models with different values of 
the o/p ratio of \HTDP \ (o/p--\HTDP ), from 0.03 (bottom 
curves) to 2.0 (top curves). As discussed by \citet{fpw04}, 
the o/p ratio is a sensitive function of the o/p H$_2$ ratio and, 
ultimately, of the gas temperature (see their Fig.~6) 
and at $T_{\rm gas}$ $<$ $\sim$15~K, it changes from $\simeq$0.03 to 
values probably larger than one (this last statement is valid if the 
curve in Fig.~6 of \citet{fpw04} is simply extrapolated at 
temperatures lower than 9~K, the minimum value in the figure\footnotemark{}
\footnotetext{At a density of 2$\times$10$^6$ cm$^{-3}$, temperature $T_{\rm gas}$
= 8~K, grain size 0.1 $\mu$m, the o/p--\HTDP \ is about 1.2 (Pineau des For\^ets, priv. 
comm.)}). In all curves, the $t_5$ models show a slight decrease in both the 
\HTDP \ column density and in the deuterium fractionation. In fact, at such high central densities: (i) 
\HTDP \ is efficiently converted into \DTHP \ and D$_3^+$ , thus lowering the total \HTDP \ column 
along the line of sight; (ii) \MOLN \  significantly freezes onto dust grains, 
so that the \NTDP /\NTHP column density ratio -- our measure of
the observed $R_{\rm D}$ -- traces regions away from the
center, where the density is lower and the temperature is
higher. If a protostar is present, the \HTDP \ column density increases again
because of the less efficient transformation of \HTHP \ into \DTHREEP , while 
$R_{\rm D}$ decreases (see also Fig.~\ref{fdeuteration}) because of the 
less abundant \DTHREEP .  From the comparison between the data and the model, one is tempted 
to conclude that indeed variations in the o/p--\HTDP \ ratio (and 
ultimately in the gas temperature of the central few thousand AU of 
dense cores) can explain the observed scatter among the cores. It is 
interesting to note that the two pre-stellar cores with high values of $R_{\rm D}$ and 
relatively low ortho--\HTDP \ column densities are both embedded in the Ophiuchus 
Molecular Cloud Complex: Oph~D and 16293E. Chemical abundances observed 
in these two cores are consistent with a lower \HTDP \ o/p ratio,
suggesting possible (slightly) larger kinetic temperatures.   

\begin{figure}
\centering
\includegraphics[width=0.4\textwidth]{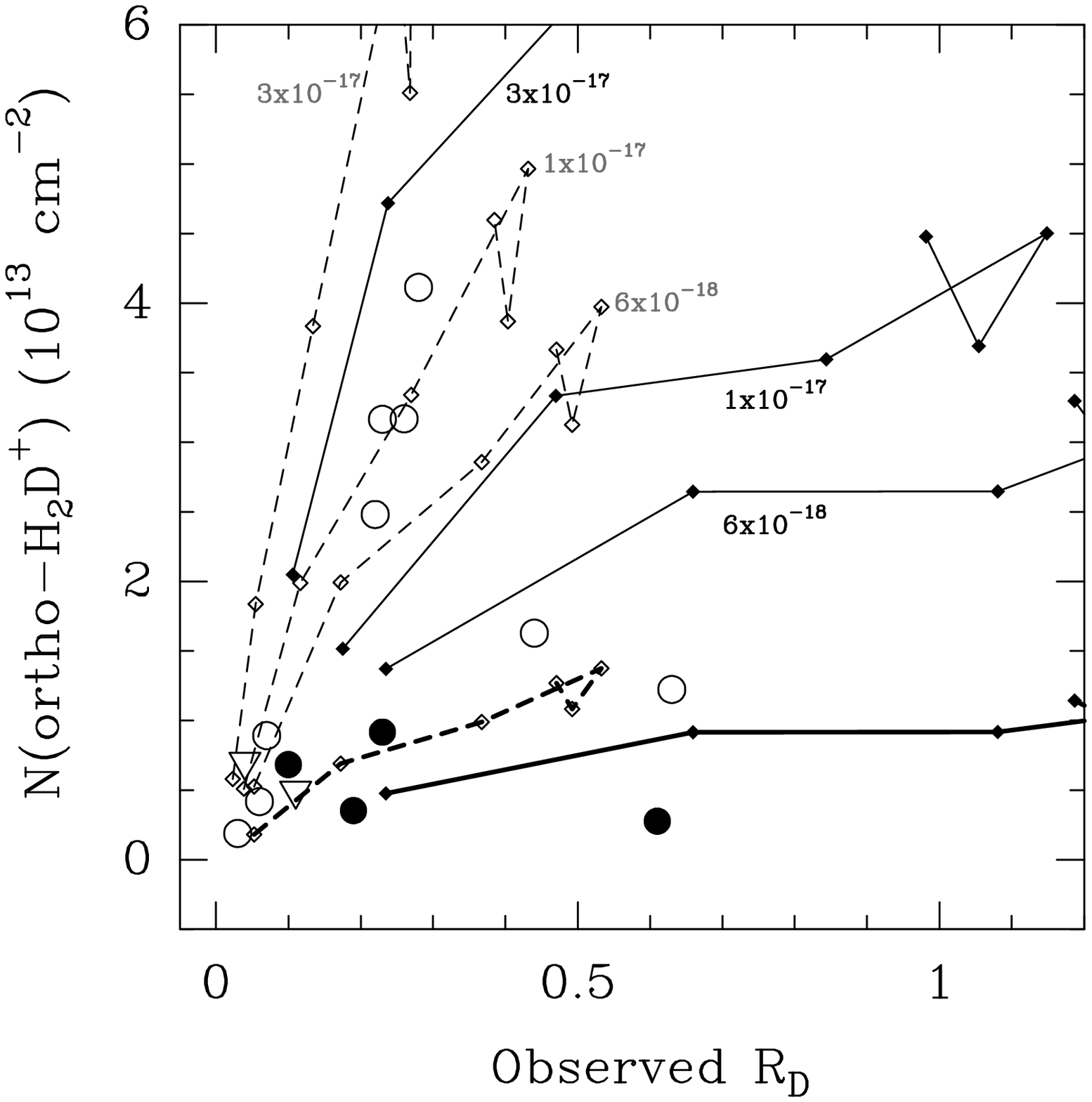}
\quad
\includegraphics[width=0.41\textwidth]{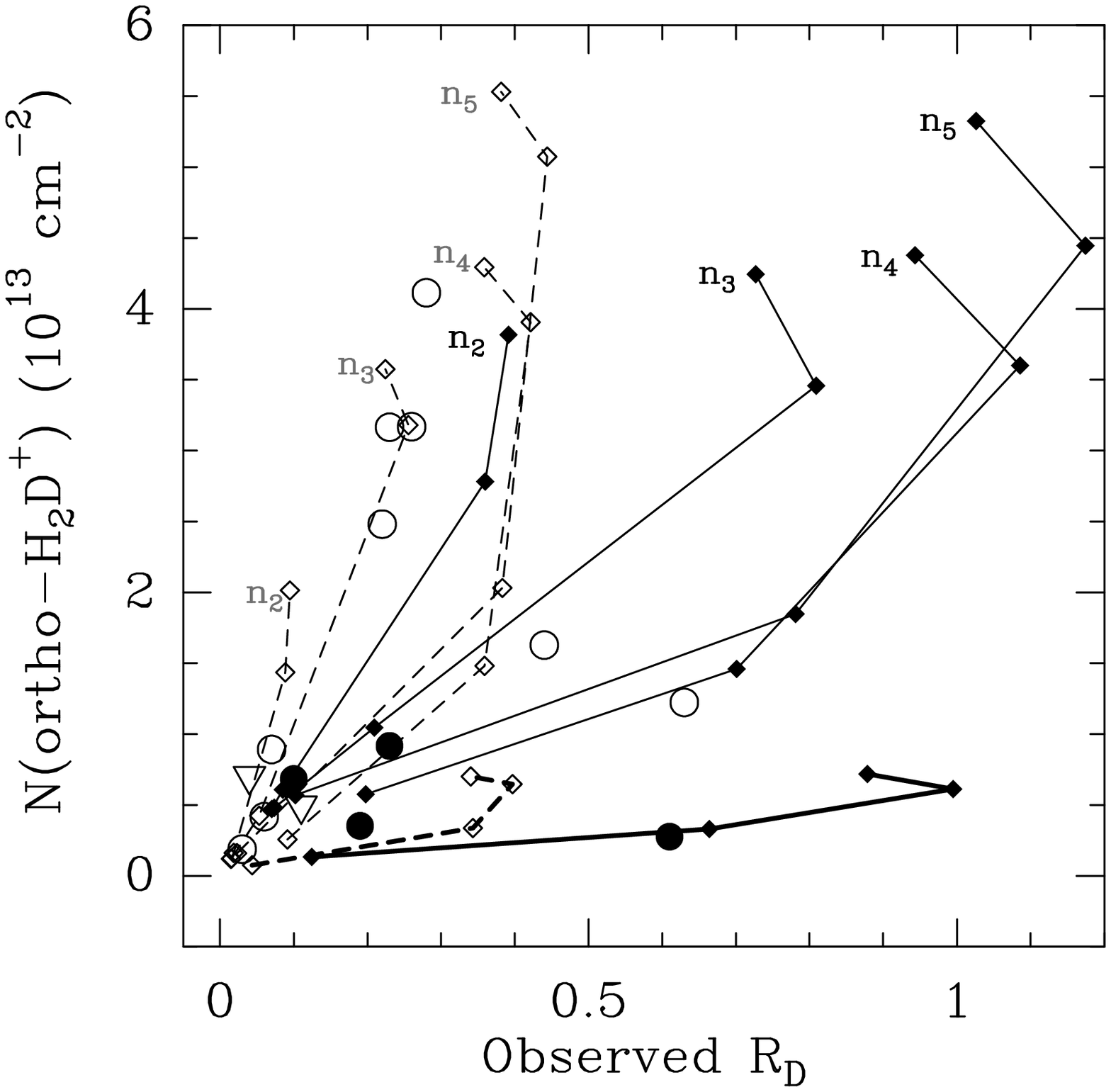}
\caption{({\it Left panel})
Same models as in the right panel of Fig.~\ref{fchem01}
but with a fixed value o/p--\HTDP \ (=0.5), and various cosmic--ray
ionization rates (from 6 to 30 $\times$10$^{-18}$ s$^{-1}$).  The 
thick dashed curve connects model results using GHR02 rates, 
$\zeta$ = 6$\times$10$^{-18}$~s$^{-1}$ and o/p--\HTDP = 0.3. This is 
an attempt to reproduce the low ortho--\HTDP \ 
column density values observed toward the $R_{\rm D}$--rich protostellar 
cores and the two Ophiuchus pre-stellar cores Oph~D and 16293E.  
({\it Right panel}) Plot showing predictions of models with
density structures equal to the \citet{cb00} cloud at times
$t_i$ ($i$ = 2, 3, 4 and 5) and corresponding central densities $n_{ci} \equiv n_i$ (see text) at different stages of chemical evolution (at $t$ = 10$^3$, 10$^4$, 10$^5$,
and 10$^6$ yr, from bottom left to top right). The o/p--\HTDP \ 
ratio has been fixed at 0.5.  The thick curves 
refer to the model of a protostellar envelope at different evolutionary 
times, where o/p--\HTDP = 0.1, to simulate a possible (slight) increase 
of the gas temperature due to the central heating source.}
\label{fchem23}
\end{figure}

Fig.~\ref{fchem23} shows two other attempts to interpret the data. 
The left panel considers exactly the same models as in Fig.~\ref{fchem01}
but now the o/p--\HTDP ratio is fixed at 0.5 (consistent with dense gas at 
9~K; see \citealt{fpw04}), whereas the free parameter
is the cosmic--ray ionization rate $\zeta$, which is varied from 6$\times$10$^{-18}$ s$^{-1}$
(bottom dashed and solid curves) and 3$\times$10$^{-17}$~s$^{-1}$ (top curves). 
We note that variations in the cosmic--ray ionization rate are known to exist in the
Galaxy \citep{vbs06}. 
The four pre--stellar cores with the largest $N(ortho-\HTDP )$ values (L~492, L~1544, 
L~694--2, and L~183) can all be reproduced by $t_2$--$t_4$  ($t_1$--$t_2$) models with the 
GHR02 (standard) rates and $\zeta$ = 1$\times$10$^{-17}$~s$^{-1}$ (with L~429 (L~183) being the most (least) dynamically evolved).   Significantly lower values of $\zeta$ ($<$ 6$\times$10$^{-18}$ s$^{-1}$) appear 
to be required in the protostellar and Ophiuchus cores. However, an alternative way to reproduce these data, without requiring extremely low $\zeta$ values,  is to lower the o/p--\HTDP \  ratio, as found before (the thick dashed curves represent models with o/p--\HTDP \ = 0.3 and $\zeta$ = 6$\times$10$^{-18}$ s$^{-1}$). In B~68, only the low o/p--\HTDP \ model curve (at early 
evolutionary times) can reproduce the low $ortho-\HTDP $ column densities 
and deuterium fractionations, the lowest in the sample.  

In the right panel of Fig.~\ref{fchem23}, we consider  five clouds  
with structure as in the $t_i$ ($i$= 2, ..., 5) models (see eqn.~\ref{eden})  and for each of 
them we follow the chemical evolution, checking the results after  
10$^3$, 10$^4$, 10$^5$, and 10$^6$ yr.  This allows us to explore
how different chemical ages can change the gas composition and avoid 
the problem of fixing the (unknown) chemical evolution time 
as in the right panel of Fig.~\ref{fchem01}.  As for the left panel,
the o/p--\HTDP \ ratio has been fixed at 0.5, except for the thick 
curves, representing the $t_{\rm 5a}$ protostellar cloud models, where 
o/p = 0.1, assuming that the gas has been (slightly) heated compared 
to the pre--stellar cores. The cosmic--ray ionization rate is fixed 
at 1.3$\times$10$^{-17}$ s$^{-1}$.  The four 
ortho-\HTDP -- rich pre-stellar cores can be reproduced by $t_3$ models, with (chemical)
ages between 10$^4$ and 10$^6$ yr, when GHR02 rate coefficients 
are used. On the other hand, the protostellar cores and the two Ophiuchus 
cores are better matched by the more dynamically evolved (centrally concentrated)
$t_{\rm 5a}$ and $t_5$ models, respectively, after about 10$^3$--10$^4$ years of (chemical)
evolution. 

\subsubsection{$N(ortho-\HTDP )$ vs. the observed $f_{\rm D}$} 
\label{s_fd}

From the models described in the previous subsection, one can 
directly derive the abundance of CO within each cloud model as a function 
of cloud radius and, as before, $N({\rm CO})$
is obtained, after integrating the CO number density along the 
line of sight and smoothing the results with a 22\arcsec \ beam width 
(to simulate observations carried out at the IRAM~30m antenna, where
most of the cores have been observed). From the model column density, 
the CO depletion factor, $f_{\rm D}$(CO) is easily calculated using the 
expression:
\begin{eqnarray}
f_{\rm D}({\rm CO}) & = & \frac{x_{\rm can}({\rm CO})}
{N({\rm CO})/N(\MOLH )} ,
\end{eqnarray}
where $x_{\rm can}$(CO) is the ``canonical'' or ``undepleted'' 
abundance of CO (assumed here equal to 9.5$\times$10$^{-5}$, 
following \citealt{flw82}, but known to be uncertain by a
factor of about 2, see e.g. \citealt{lkg94}).  

\begin{figure}
\centering
\includegraphics[width=0.4\textwidth]{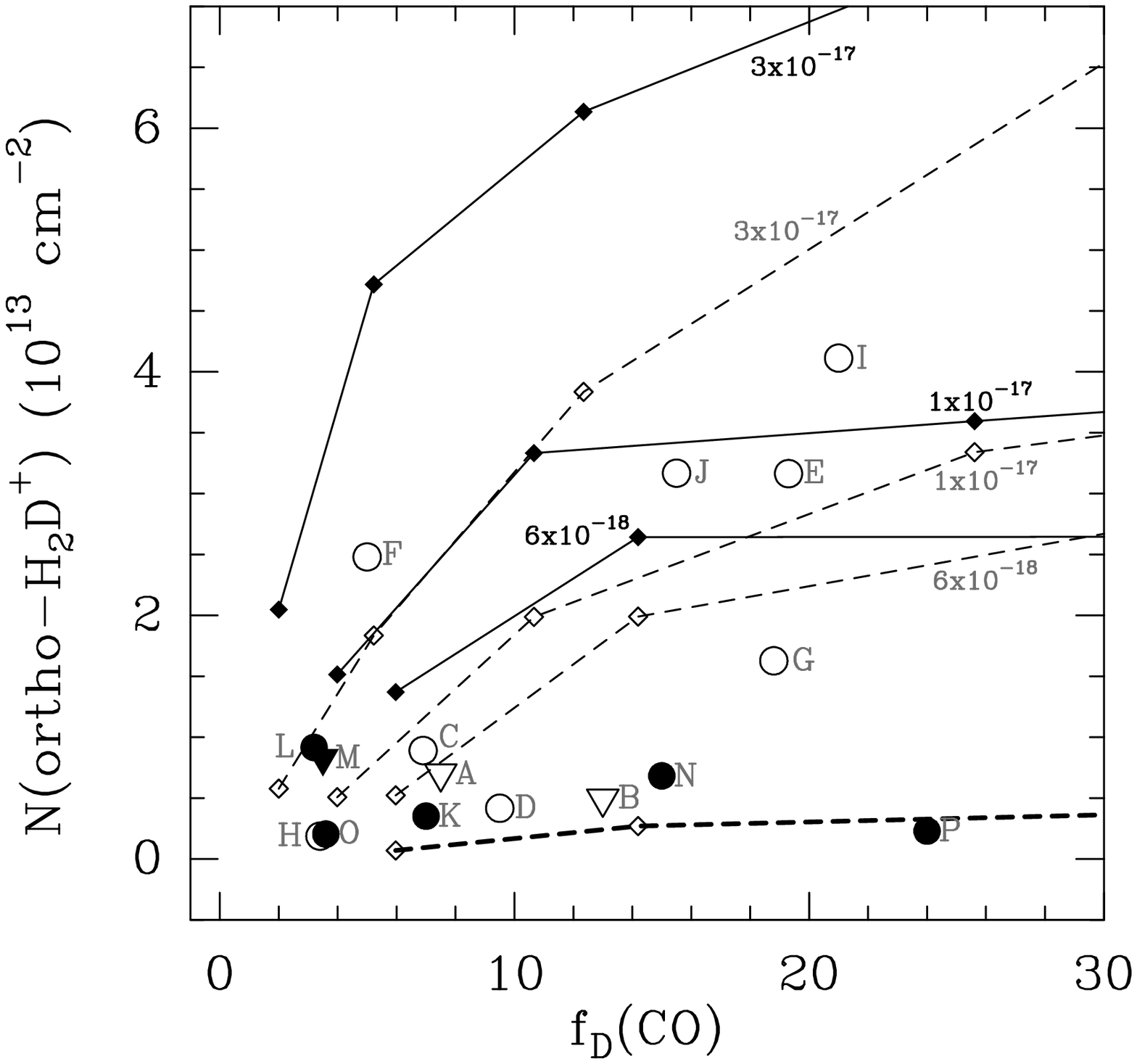}
\quad
\includegraphics[width=0.41\textwidth]{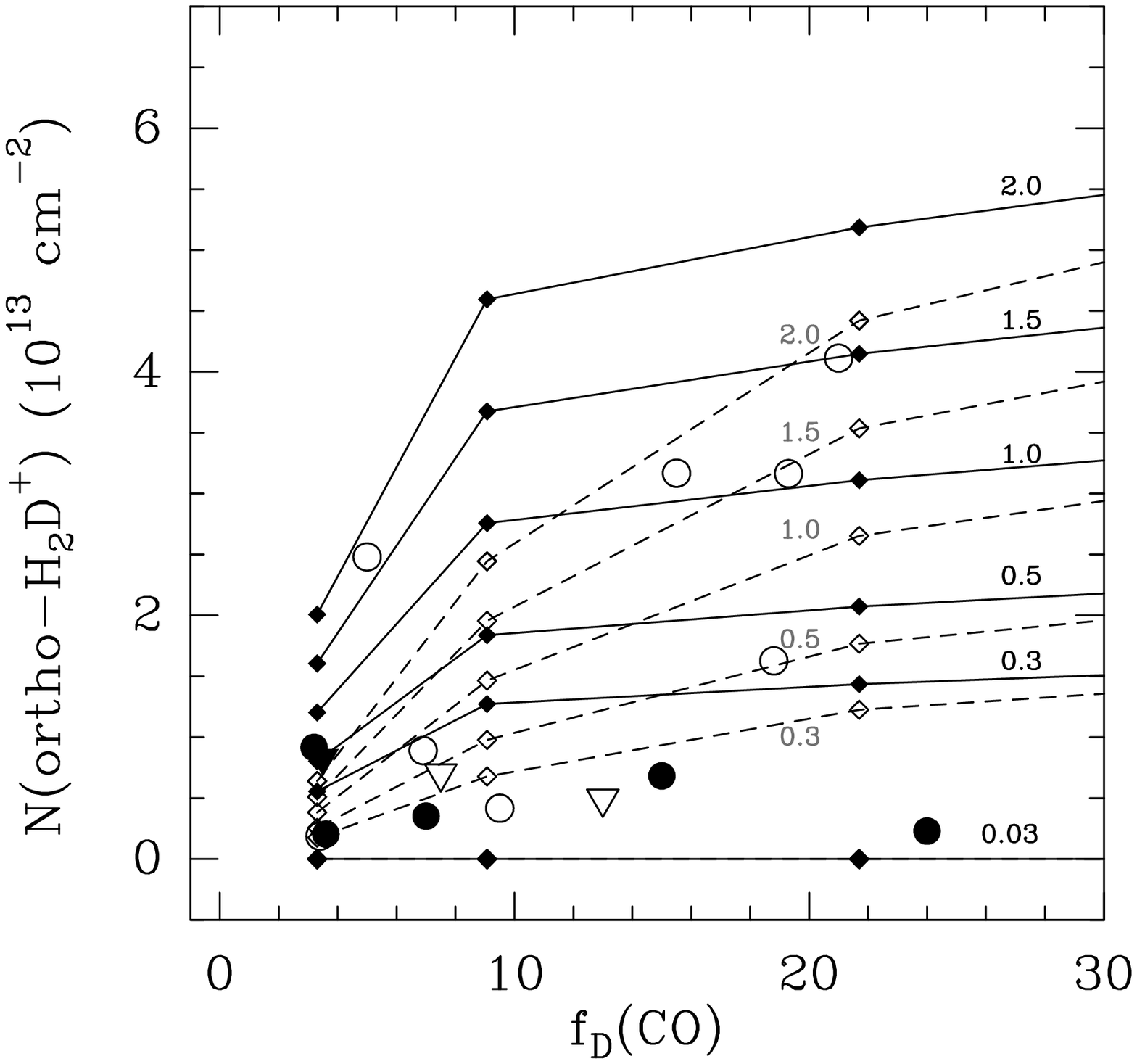}
\caption{$N(ortho-\HTDP )$ vs the observed CO depletion factor,
$f_{\rm D}$(CO). The same models $t_i$ (i=1, ..., 5) described for the
previous figures are used here, varying the cosmic--ray ionization
rate $\zeta$ ({\it left panel}) and the o/p--\HTDP \ ratio ({\it right panel})
within the range of Fig.~\ref{fchem01} and \ref{fchem23}. The thick dashed curve refer to (GHR02) models with o/p-\HTDP \ ratio of 0.1 and $\zeta$ = 6$\times$10$^{-18}$ s$^{-1}$. 
As found before, variations in the o/p-\HTDP \ ratio, plus differences in the physical structure, 
provide a way to reproduce the whole observed spread in the data. The labels in the left panel 
correspond to the source names: A$\equiv$L~1498, B$\equiv $TMC--2,
C$\equiv $TMC--1C, D$\equiv $L~1517B, E$\equiv $L~1544, F$\equiv $L~183,
G$\equiv $Oph~D, H$\equiv $B~68, I$\equiv $L~429, J$\equiv $L~694-2,
K$\equiv $NGC~1333--\DCOP , L$\equiv $B~1, M$\equiv $IRAM~04191,
N$\equiv $L~1521F, O$\equiv $Ori~B9.}
\label{fdep}
\end{figure}

To compare our model predictions with the data, we collect from the 
literature the values of observed $f_{\rm D}$(CO) and plotted them 
versus $N(ortho-\HTDP )$ in Fig.~\ref{fdep}.  The majority of
the $f_{\rm D}$(CO) data comes from \citet{ccw05}, except for:
(i) TMC--1C ($f_{\rm D}$(CO) = 6.9, from \citealt{scg07}), (ii) L~183 ($f_{\rm D}$ = 5, from \citealt{ppa05}); 
(iii) NGC~1333--\DCOP \ ($f_{\rm D}$ = 12, from \citealt{jsv02}); 
(iv) B~1 ($f_{\rm D}$(CO) = 3.2, from \citealt{lrg02});
(v) IRAM~04191 ($f_{\rm D}$(CO) = 3.5 from \citealt{ba04});
(vi) OriB~9 ($f_{\rm D}$(CO) = 3.6, from \citealt{hhl06}, for $N(\MOLH )$,
and \citealt{cm95}, for $N(\rm CO)$). Also, for L~1544, L~429, L~694-2 and Oph~D, the new values of $N(\MOLH )$, adopted to take into account the temperature structure (see Tab.~\ref{tcolumn}), imply different values of $f_{\rm D}$ (larger by a factor of about 1.4, the ratio between the new and old $N(\MOLH )$ values, as explained in Sect.~\ref{scolumn}) from what reported by \cite{ccw05}. The data and model results are shown in Fig.~\ref{fdep}. In general, the presence of embedded {\it young} stellar objects appears to lower the \HTDP \ column density, without affecting much the amount of CO freeze-out, which is probably still large in the high density and cold protostellar envelopes. 
The possible (small) temperature increase caused by the central heating, is thus not sufficient to significantly release CO back in the gas phase, but it can affect the o/p--\HTDP \ ratio (and, consequently, the ortho-\HTDP \ column density), as discussed in the previous sub--section. 

In the left panel of Fig.~\ref{fdep}, the data are compared with the 
same models described in Fig.~\ref{fchem23} (left panel; $t_i$, $i$=1, 2, ...5),
where the cosmic--ray ionization rate is varied from 
6$\times$10$^{-18}$ s$^{-1}$ to 3$\times$10$^{-17}$ s$^{-1}$.
The o/p--\HTDP \ ratio has been fixed at 0.5, except for the thick 
dashed curve, where o/p = 0.1, value probably more appropriate 
for protostellar cores (see previous sections).  Now, 
$t_2$--$t_3$ models (depending on the rate coefficient values adopted
for the proton--deuteron exchange reactions), with $\zeta$ $\simeq$
10$^{-17}$ s$^{-1}$, can explain the observations toward the 
ortho-\HTDP -- rich pre--stellar cores, with the exception of L~183, where the low $f_{\rm D}$ 
value suggests a younger dynamical phase, consistent with what found in the previous sub-section (we also note that  L~183 and TMC-1C appear to have similar ages,  but with different cosmic ray ionization rate and/or o/p-\HTDP \ ratio, see right panel of Fig.~\ref{fdep}).   Lower o/p--\HTDP \ ratios are needed to reproduce the protostellar and Ophiuchus cores, as found for the 
$N(ortho-\HTDP )$ vs. $R_{\rm D}$ relation. 

The right panel of Fig.~\ref{fdep} shows the same set of data and 
models, but now $\zeta$ has been fixed at 1.3$\times$10$^{-17}$ s$^{-1}$,
whereas the o/p--\HTDP \ ratio has been changed as in Fig.~\ref{fchem01}.
As already noted, L~1544, L~429, L~694-2, and L~183 data are best reproduced 
by large values of the o/p--\HTDP \ ratio, and the appropriate 
physical structure is that of $t_1$--$t_3$ models , similarly to the left
panel (with L~183 being the least evolved). This implies cores slightly more evolved than what found in Fig.~\ref{fchem01} (right panel) and Fig.~\ref{fchem23} (left panel), where (standard rate) models between $t_1$ and $t_2$ were preferred. The small discrepancy can be understood if the predicted CO depletion factor is  too large compared with observations. Reasons for this could be: 
(i) we are missing an important desorption mechanism (besides the cosmic--ray impulsive 
heating and the thermal desorption, the latter not being efficient at 
the temperatures of these objects); (ii) unlike our spherical and isolated model cores, real 
cloud cores are embedded in molecular clouds where CO is
quite abundant.  Thus, the observed ``extra'' gaseous CO may 
be part of the undepleted material accreting onto the core 
from the surrounding molecular cloud (see also \citealt{scg07} for a similar
conclusion in the particular case of TMC--1C). 

\section{Conclusions}
\label{scon}

Low--mass dense cloud cores have been observed with the CSO antenna 
at the frequency of the ortho--\HTDP (1$_{1,0}$--1$_{1,1}$) line.  The 
main conclusions of this work are:

\noindent
1)  In starless cores, the line has been detected in 7 (out of 10) objects. 
The brightest lines ($T_{\rm mb}$ = 0.7-0.9~K) are observed toward the densest
and more dynamically evolved starless cores (L~1544, L~183, Oph~D, L~429, and L~694-2), 
where the deuterium fractionation and the CO depletion factor are largest. 
Non-detections are found in cores less centrally concentrated and dense than the
rest of the sample (L~1498, TMC--2, and B~68). In B~68, recent APEX observations
have detected a faint ortho--\HTDP \ line with intensities consistent with 
our upper limit. 

\noindent
2) Significant differences in line widths are observed also among 
starless cores, with the narrowest lines ($\Delta {\rm v}$ $\simeq$ 0.4~\kms ) 
found in TMC--1C, L~1517B, Oph~D, and L~183, whereas L~1544, L~429, and L~694-2 
show $\Delta {\rm v}$ = 0.5--0.7~\kms.  

\noindent
3) The ortho--\HTDP \trans \ line has been detected in 4 out of 6 protostellar cores, 
where we find $T_{\rm mb}$ $\leq$ 0.5~K (the largest value observed is toward L~1521F, 
which hosts a very low luminosity object recently detected by Spitzer).  
The broadest lines are observed toward the two cores in Perseus NGC~1333--\DCOP \
and B~1, where $\Delta {\rm v}$ $\simeq$1~\kms. The ortho--\HTDP \trans \ lines have broader 
non--thermal widths than \NTDP (2--1) lines, even in L~1544, where the two 
transitions have similar extension and morphology. 

\noindent
4) The ortho--\HTDP \ column densities range between 2 and 
40$\times$10$^{12}$~cm$^{-2}$ in starless cores and between 
2 and 9 $\times$10$^{12}$~cm$^{-2}$ in protostellar cores. 
Thus, protostars in the earliest stages of their evolution 
appear to have already changed the chemical structure of their envelopes, by 
lowering the ortho--\HTDP \ fractional abundance but not their 
deuterium fractionation. This is probably due to (a few degree) variation
of the gas temperature, which strongly affects the o/p--\HTDP \ ratio. 
A similar effect is also found in the two Ophiuchus cores, suggesting 
a (slightly) larger gas temperature than in the other (mainly Taurus 
and Perseus) cores. 

\noindent
5) Simple models suggest that variations in the gas kinetic temperature, 
CO depletion factor, grain size distribution, cosmic--ray ionization 
rate and volume densities can largely affect the \HTDP \ abundance 
relative to \HTHP . In particular, gas temperatures above 15~K, low 
CO depletion factors and large abundance of negatively charged small 
dust grains or PAHs drastically reduce the 
deuterium fractionation to values inconsistent with 
those observed toward pre--stellar and protostellar cores.  

\noindent
6) Plots of the ortho--\HTDP \ column density vs. (i) the  
deuterium fractionation observed in species such as \NTHP \ and \AMM , 
and (ii) the observed CO depletion factor, show a large scatter.  
The data can be reproduced by chemical models of
centrally concentrated cores assuming variations in the o/p
ratio of \HTDP \ (ultimately linked to kinetic temperature variations). 
Changes in the cosmic--ray ionization rate between 6 and 
30$\times$10$^{-18}$s$^{-1}$ can also explain part of the scatter, but 
not those objects with large deuterium fractionations {\it and} 
low ortho--\HTDP \ column densities
(such as the Ophiuchus pre--stellar cores and the protostellar ones), for 
which a decrease of the o/p--\HTDP \ ratio is required.  
The most deuterated and \HTDP--rich 
objects are better reproduced by chemical models of centrally concentrated 
cores with $n_c$ = a few $\times$ 10$^6$ \percc \ and 
chemical ages between 10$^4$ and 10$^6$~yr. Protostellar cores, plus the two Ophiuchus cores, are better matched by  lower o/p-\HTDP \ ratios, more dynamically evolved (central densities $\ge$10$^7$ cm$^{-3}$) models and chemical ages of $\simeq$10$^3$-10$^4$ years. To better constrain dynamical and chemical ages, the rate coefficient for the proton-deuteron exchange reaction needs to be well defined. 

\noindent
7) The para-\HTHOP (1$_1^-$--2$_1^+$) upper limits are consistent 
with radiative transfer calculations if the fractional abundance of 
\HTHOP \ is $\la$10$^{-8}$.  The para--\DTHP (1$_{1,0}$--1$_{0,1}$) 
upper limits provide an upper limit of the para--\DTHP \ to 
ortho--\HTDP \ column density ratio ($\equiv p/o$).  We find 
$p/o$$<$1 in all the sources (except for L~183 and NGC~1333--\DCOP , where $p/o < 3$), 
consistent with  chemical model predictions of high density 
(2$\times$10$^6$~\percc) and low temperature ($T_{\rm kin} < 10$~K) clouds
(\citealt{fpw04}). 

More accurate determinations of temperature and density profiles, as well 
as observations of the para--\HTDP (1$_{0,1}$-1$_{0,0}$) line at 
1.37~THz, are sorely needed to put more stringent constraints on gas--grain 
chemical processes in dense cloud cores.

\appendix

\section{The search of para-\HTHOP }

The \HTHOP \ line was observed to investigate the chemistry
of oxygen in dense cores and, with the help of chemical models,  to put some
constraints on the oxygen abundance, which, analogously to CO, significantly
affects the deuterium fractionation.

We searched for para-\HTHOP (1$^-_1$--2$^+_1$) in seven dense cores but
only upper limits were measured. Tab.~\ref{thtop} lists the results of
this search, including the rms noise and the corresponding upper limits
of the column density, which have been calculated in two different ways:
(i) using the volume density and kinetic temperature values listed in
Tab.~\ref{tcolumn} ($N_1$, column 6), and (ii) assuming a volume density
of 10$^5$ \percc \ and $T_{\rm kin}$ = 10~K ($N_2$, column 8).  In both
cases, the RADEX\footnotemark{}
\footnotetext{The RADEX code is available on-line at the URL
http://www.sron.rug.nl/~vdtak/radex/radex.php.} code has been used 
(\citealt{vbs07}). The observed upper limits (see Tab.~\ref{thtop}) are compatible with 
para--\HTHOP \ fractional abundance upper limits of 10$^{-8}$ according to calculation 
made using the Ratran code \citep{hv00}.

\begin{table}
\begin{center}
\caption{ p--\HTHOP \ column density upper limits. \label{thtop}}
\begin{tabular}{lcccccccc}
\hline\hline
Source &  $T_{\rm rms}$ & $T_{\rm R}$ & $\Delta {\rm v}^{(a)}$ & $\eta_b$  & $N_1^{(b)}$ &
$T_{\rm ex_1}$ & $N_2^{(c)}$ & $T_{\rm ex_2}$   \\
Name   & (K) & (K) & (\kms ) &  & (\cmsq ) & (K) & (10$^{15}$ \cmsq ) & (K) \\
\hline
IRAM~04191 & 0.054 & $<$0.16 & 0.35 & 0.65  & $<$3.5 & 7.0 & $<$1.1 & 4.1 \\
L~1521F    & 0.042 & $<$0.13 & 0.31 & 0.65  &  $<$0.90 & 6.1 & $<$0.85 & 4.1 \\
L~1544     & 0.025 & $<$0.08 & 0.33 & 0.65 &  $<$0.55 & 6.1 & $<$0.63 & 3.9 \\
L~183      & 0.031 & $<$0.09 & 0.23 & 0.6 & $<$0.53 & 5.9 & $<$0.50 & 3.9 \\
Oph~D    & 0.083 & $<$0.25 & 0.24 & 0.6 & $<$0.30 & 6.0 & $<$1.1 & 4.3 \\
L~429      & 0.043 & $<$0.13 & 0.40 & 0.6 & $<$1.5 & 5.5 & $<$1.1 & 4.1 \\
L~694-2   & 0.046 & $<$0.14 & 0.28 & 0.6 & $<$0.90 & 5.9 & $<$0.80 & 4.1 \\
\hline
\end{tabular}
\begin{list}{}{}
\item[$^{\mathrm{a}}$] Assumed value, derived from previous \NTHP (1--0) observations (see text).
\item[$^{\mathrm{b}}$] Column density value obtained using $n(\MOLH )$ and $T_{\rm kin}$ from Tab.~\ref{tcolumn}.
\item[$^{\mathrm{c}}$] Column density value assuming $n(\MOLH )$ = 10$^5$ \percc \ and $T_{\rm kin}$ = 10~K.
\end{list}
\end{center}
\end{table}

%Fig.~\ref{fh3op} shows predictions of the para-\HTHOP \ line in a dense core
%with the same physical characteristics as L~1544, for different values of the
%\HTHOP \ abundance, using the Ratran code \citep{hv00}.
%The green histograms show the model line profile, taking into account the
%dust emission at 307 GHz, whereas the red histogram does not consider
%dust emission. The comparison between observed and synthetic line intensities
%allows us to conclude that our results (Table~\ref{thtop}) are consistent with
%para--\HTHOP \ fractional abundance upper limits of 10$^{-8}$.

%\begin{figure}
%\includegraphics[angle=-90,scale=.55]{h3o+_model.ps}
%\caption{Radiative transfer simulation of the expected intensity of the
%307~GHz \HTHOP \ line in an object similar to L~1544, for different
%values of the para--\HTHOP \ abundance.  Dust emission is (not) taken
%into account in the green (red) histograms.}
%\label{fh3op}
%\end{figure}

\begin{acknowledgements}
We thank an anonymous referee for his/her very detailed review, which greatly improved the paper. 
PC acknowledges support from the 
Italian Ministry of Research and University within a PRIN
project. Part of this research has been supported by NSF grant AST 05-40882 to the CSO. 
The authors thank the staff of the CSO telescope for their support. We also thank E. Hugo and S. Schlemmer for providing 
their collisional rates prior to publication.
\end{acknowledgements}


\begin{thebibliography}{}

\bibitem[Aikawa et al. (2001)]{aoi01} Aikawa, Y., Ohashi, N., Inutsuka, S.-i., Herbst, E., \& Takakuwa, S., 2001, \apj, 552, 639 

\bibitem[Aikawa et al. (2005)]{ahr05} Aikawa, Y., Herbst, E., Roberts, H., \& Caselli, P., 2005, \apj, 620, 330

\bibitem[Aikawa et al.(2008)]{awg08} Aikawa, Y., Wakelam, V., Garrod, R.~T., \& Herbst, E., 2008, \apj, 674, 984 

\bibitem[Alves et al. (1998)]{all98} Alves, J., Lada, C.~J., Lada, E.~A., Kenyon, S.~J., \& Phelps, R., 1998, \apj, 506, 292 

\bibitem[Alves et al. (2001)]{all01} Alves, J., Lada, C.~J., \& Lada, E.~A., 2001, \nat, 409, 159 

\bibitem[Amano \& Hirao (2005)]{ah05} Amano, T., \& Hirao, T., 2005, Journal of Molecular Spectroscopy, 233, 7

\bibitem[Andr{\'e} et al. (2000)]{awb00} Andr{\'e}, P., Ward-Thompson, D., \& Barsony, M.\ 2000, Protostars and Planets IV, 59 

\bibitem[Bacmann et al. (2002)]{blc02} Bacmann, A., Lefloch, B., Ceccarelli, C., et al., 2002, \aap, 389, L6 

\bibitem[Bacmann et al. (2003)]{blc03} Bacmann, A., Lefloch, B., Ceccarelli, C., et al., 2003, \apjl, 585, L55

\bibitem[Belloche et al. (2002)]{bad02} Belloche, A., Andr{\'e}, P., Despois, D., \& Blinder, S., 2002, \aap, 393, 927 

\bibitem[Belloche \& Andr{\'e} (2004)]{ba04} Belloche, A., \& Andr{\'e}, P., 2004, \aap, 419, L35

\bibitem[Benson \& Myers (1989)]{bm89} Benson, P.~J., \& Myers, P.~C., 1989, \apjs, 71, 89

\bibitem[Benson et al. (1998)]{bcm98} Benson, P.~J., Caselli, P., \& Myers, P.~C., 1998, \apj, 506, 743 

\bibitem[Bergin et al. (2001)]{bcl01} Bergin, E.~A., Ciardi, D.~R., Lada, C.~J., Alves, J., \& Lada, E.~A., 2001, \apj, 557, 209 

\bibitem[Bergin \& Snell (2002)]{bs02} Bergin, E.~A., \& Snell, R.~L., 2002, \apjl, 581, L105

\bibitem[Bergin et al. (2002)]{bah02} Bergin, E.~A., Alves, J., Huard, T., \& Lada, C.~J., 2002, \apjl, 570, L101 

\bibitem[Bergin et al. (2006)]{bmv06} Bergin, E.~A., Maret, S., van der Tak, F.~F.~S., et al., 2006, \apj, 645, 369 

\bibitem[Bergin \& Tafalla(2007)]{bt07} Bergin, E.~A., \& Tafalla, M., 2007, \araa, 45, 339 

\bibitem[Bianchi et al. (2003)]{bga03} Bianchi, S., Gon{\c c}alves, J., Albrecht, M., et al., 2003, \aap, 399, L43

\bibitem[Bisschop et al. (2006)]{bfo06} Bisschop, S.~E., Fraser, H.~J., {\"O}berg, K.~I., van Dishoeck, E.~F., \& Schlemmer, S., 2006, \aap, 449, 1297 

\bibitem[Black et al.(1990)]{bvw90} Black, J.~H., van Dishoek, E.~F., Willner, S.~P., \& Woods, R.~C., 1990, \apj, 358, 459 

\bibitem[Bonnor (1956)]{b56} Bonnor, W.~B., 1956, \mnras, 116, 351 

\bibitem[Bottinelli et al. (2004a)]{bcl04} Bottinelli, S., Ceccarelli, C., Lefloch, B. et al., 2004a, ApJ 615, 354

\bibitem[Bottinelli et al. (2004b)]{bcn04} Bottinelli, S., Ceccarelli, C., Neri, R., et al., 2004b, ApJ 617, L69

\bibitem[Bottinelli et al. (2007)]{bcw07} Bottinelli, S., Ceccarelli, C., Williams, J.~P., \& Lefloch, B., 2007, \aap, 463, 601 

\bibitem[Bourke et al. (2006)]{bme06} Bourke, T. L., Myers, P. C., Evans, N. J., II, et al., 2006, ApJ 649, L37

\bibitem[Butner et al. (1995)]{bll95} Butner, H.~M., Lada, E.~A., \& Loren, R.~B., 1995, \apj, 448, 207

\bibitem[Caselli \& Myers (1995)]{cm95} Caselli, P., \&Myers, P.~C., 1995, \apj, 446, 665

\bibitem[Caselli et al. (1999)]{cwt99} Caselli, P., Walmsley, C.~M., Tafalla, M., Dore, L., \& Myers, P.~C., 1999, \apjl, 523, L165 

\bibitem[Caselli et al. (2002a)]{cwz02a} Caselli, P., Walmsley, C.~M., Zucconi, A., Tafalla, M., Dore, L., \& Myers, P.~C., 2002a, \apj, 565, 331 

\bibitem[Caselli et al. (2002b)]{cwz02b} Caselli, P., Walmsley, C.~M., Zucconi, A., Tafalla, M., Dore, L., \& Myers, P.~C., 2002b, \apj, 565, 344 

\bibitem[Caselli et al. (2002c)]{cbm02} Caselli, P., Benson, P.~J., Myers, P.~C., \& Tafalla, M., 2002c, \apj, 572, 238 

\bibitem[Caselli et al. (2003)]{cvc03} Caselli, P., van der Tak, F.~F.~S., Ceccarelli, C., \& Bacmann, A., 2003, \aap, 403, L37 

\bibitem[Caux et al. (1999)]{ccc99} Caux, E., Ceccarelli, C., Castets, A., et al., 1999, A\&A 347, L1

\bibitem[Cazaux et al. (2003)]{ctc03} Cazaux, S., Tielens, A.~G.~G.~M., Ceccarelli, C., et al., 2003, \apjl, 593, L51

\bibitem[Ceccarelli et al. (1998)]{ccl98} Ceccarelli, C., Castets, A., Loinard, L., Caux, E., \& Tielens, A.~G.~G.~M., 1998, \aap, 338, L43

\bibitem[Ceccarelli \& Dominik(2005)]{cd05} Ceccarelli, C., \& Dominik, C., 2005, \aap, 440, 583 (CD05)

\bibitem[Ceccarelli et al.(2007)]{cch07} Ceccarelli, C., Caselli, P., Herbst, E., Tielens, A.~G.~G.~M., \& Caux, E., 2007, Protostars and Planets V, 47 

\bibitem[Charnley et al.(1997)]{ctr97} Charnley, S.~B., Tielens, A.~G.~G.~M., \& Rodgers, S.~D.,1997, \apjl, 482, L203 

\bibitem[Ciolek \& Basu (2000)]{cb00} Ciolek, G.~E., \& Basu, S., 2000, \apj, 529, 925 

\bibitem[Crapsi et al. (2004)]{ccw04} Crapsi, A., Caselli, P., Walmsley, C.~M., et al., 2004, \aap, 420, 957 

\bibitem[Crapsi et al. (2005)]{ccw05} Crapsi, A., Caselli, P., Walmsley, C.~M., et al., 2005, \apj, 619, 379 

\bibitem[Crapsi et al. (2007)]{ccw07} Crapsi, A., Caselli, P., Walmsley, M.~C., \& Tafalla, M.\ 2007, \aap, 470, 221 

\bibitem[Dalgarno \& Lepp (1984)]{dl84} Dalgarno, A., \& Lepp, S.\ 1984, \apjl, 287, L47 

\bibitem[Dalgarno (2006)]{d06} Dalgarno, A., 2006, Proceedings of the National Academy of Science, 103, 12269 

\bibitem[Draine \& Sutin (1987)]{ds87} Draine, B.~T., \& Sutin, B., 1987, \apj, 320, 803

\bibitem[Ebert (1955)]{e55} Ebert, R., 1955, Zeitschrift fur Astrophysik, 36, 222 

\bibitem[Evans et al. (2001)]{ers01} Evans, N.~J., II, Rawlings, J.~M.~C., Shirley, Y.~L., \& Mundy, L.~G., 2001, \apj, 557, 193 

\bibitem[Flower \& Pineau des For{\^e}ts(2003)]{fp03} Flower, D.~R., \& Pineau des For{\^e}ts, G., 2003, \mnras, 343, 390 

\bibitem[Flower et al. (2004)]{fpw04} Flower, D.~R., Pineau des For{\^e}ts, G., \& Walmsley, C.~M., 2004, \aap, 427, 887 

\bibitem[Flower et al. (2005)]{fpw05} Flower, D.~R., Pineau Des For{\^e}ts, G., \& Walmsley, C.~M., 2005, \aap, 436, 933 

\bibitem[Flower et al. (2006a)]{fpw06a} Flower, D.~R., Pineau Des For{\^e}ts, G., \& Walmsley, C.~M., 2006a, \aap, 449, 621 

\bibitem[Flower et al. (2006b)]{fpw06b} Flower, D.~R., Pineau Des For{\^e}ts, G., \& Walmsley, C.~M., 2006b, \aap, 456, 215 

\bibitem[Frerking et al. (1982)]{flw82} Frerking, M.~A., Langer, W.~D., \& Wilson, R.~W., 1982, \apj, 262, 590 

\bibitem[Galli et al. (2002)]{gwg02} Galli, D., Walmsley, M., \& Gon{\c c}alves, J., 2002, \aap, 394, 275 

\bibitem[Garrod et al. (2006)]{gpc06} Garrod, R., Park, I.~H, Caselli, P.,Herbst, E., 2006, Faraday Discuss., 2006, 133, 51

\bibitem[Garrod et al. (2007)]{gwh07} Garrod, R.~T., Wakelam, V., \& Herbst, E., 2007, \aap, 467, 1103 

\bibitem[Gerin et al. (2006)]{glp06} Gerin, M., Lis, D.~C., Philipp, S., et al., 2006, \aap, 454, L63

\bibitem[Gerlich et al. (2002)]{ghr02} Gerlich, D., Herbst, E., \& Roueff, E.\ 2002, \planss, 50, 1275 (GHR02)

\bibitem[Girart et al. (2000)]{geh00} Girart, J.~M., Estalella, R., Ho, P.~T.~P., \& Rudolph, A.~L.\ 2000, \apj, 539, 763

\bibitem[Goldsmith (2001)]{g01} Goldsmith, P.~F.\ 2001, \apj, 557, 736

\bibitem[Gomez et al. (1994)]{gct94} Gomez, J.~F., Curiel, S., Torrelles, J.~M., et al., 1994, \apj, 436, 749 

\bibitem[Harju et al. (1993)]{hww93} Harju, J., Walmsley, C.~M., \& Wouterloot, J.~G.~A.\ 1993, \aaps, 98, 51 

\bibitem[Harju et al. (2006)]{hhl06} Harju, J., Haikala, L. K., Lehtinen, K. et al., 2006, A\&A 454, L55

\bibitem[Hartmann et al. (2001)]{hbb01} Hartmann, L., Ballesteros-Paredes, J., \& Bergin, E.~A., 2001, \apj, 562, 852 

\bibitem[Harvey et al. (2003)]{hwl03} Harvey, D.~W.~A., Wilner, D.~J., Lada, C.~J., Myers, P.~C., \& Alves, J.~F., 2003, \apj, 598, 1112 

\bibitem[Hasegawa et al. (1992)]{hhl92} Hasegawa, T.~I., Herbst, E., \& Leung, C.~M., 1992, \apjs, 82, 167 

\bibitem[Hasegawa \& Herbst (1993)]{hh93} Hasegawa, T.~I., \& Herbst, E., 1993, \mnras, 261, 83 

\bibitem[Hatchell (2003)]{h03} Hatchell, J., 2003, \aap, 403, L25 

\bibitem[Hatchell et al. (2005)]{hrf05} Hatchell, J., Richer, J.~S., Fuller, G.~A., et al., 2005, \aap, 440, 151

\bibitem[Hily-Blant et al.(2008)]{hwp08} Hily-Blant, P., Walmsley, M., Pineau Des For{\^e}ts, G., \& Flower, D.\ 2008, \aap, 480, L5

\bibitem[Hirano et al. (1999)]{hkm99} Hirano, N., Kamazaki, T., Mikami, H., Ohashi, N., \& Umemoto, T.\ 1999, Proceedings of Star Formation, Editor: T.~Nakamoto, Nobeyama Radio Observatory, p.~181 

\bibitem[Hogerheijde \& van der Tak (2000)]{hv00} Hogerheijde, M.~R., \& van der Tak, F.~F.~S.\ 2000, \aap, 362, 697 

\bibitem[Hogerheijde et al. (2006)]{hce06} Hogerheijde, M. R., Caselli, P., Emprechtinger, M., et al., 2006, A\&A 454, L59

\bibitem[Hotzel et al. (2002)]{hhj02} Hotzel, S., Harju, J., \& Juvela, M., 2002a, \aap, 395, L5 

\bibitem[J{\o}rgensen et al. (2002)]{jsv02} J{\o}rgensen, J.~K., Sch{\"o}ier, F.~L., \& van Dishoeck, E.~F., 2002, \aap, 389, 908 

\bibitem[Keto \& Field (2005)]{kf05} Keto, E., \& Field, G., 2005, \apj, 635, 1151 

\bibitem[Kirk et al. (2007)]{kwa07} Kirk, J.~M., Ward-Thompson, D., \& Andr{\'e}, P., 2007, \mnras, 375, 843 

\bibitem[Kuiper et al. (1996)]{klv96} Kuiper, T.~B.~H., Langer, W.~D., \& Velusamy, T., 1996, \apj, 468, 761 

\bibitem[Lacy et al. (1994)]{lkg94} Lacy, J.~H., Knacke, R., Geballe, T.~R., \& Tokunaga, A.~T., 1994, \apjl, 428, L69 

\bibitem[Lada et al. (1991)]{lbs91} Lada, E.~A., Bally, J., \& Stark, A.~A., 1991, \apj, 368, 432

\bibitem[Lada et al. (1994)]{llc94} Lada, C.~J., Lada, E.~A., Clemens, D.~P., \& Bally, J., 1994, \apj, 429, 694 

\bibitem[Lada et al. (2002)]{lba02} Lada, C.~J., Bergin, E.~A., Alves, J.~F., \& Huard, T.~L., 2002, Bulletin of the American Astronomical Society, 34, 1157 

\bibitem[Larsson et al.(1997)]{ldl97} Larsson, M., Danared, H., Larson, {\AA}., et al., 1997, Physical Review Letters, 79, 395 

\bibitem[Lee et al. (1996)]{lbh96} Lee, H.-H., Bettens, R.~P.~A., \& Herbst, E., 1996, \aaps, 119, 111 

\bibitem[Lepp \& Dalgarno (1988)]{ld88} Lepp, S., \& Dalgarno, A., 1988, \apj, 324, 553 

\bibitem[Lis et al. (2001)]{lkp01} Lis, D.~C., Keene, J., Phillips, T.~G., Schilke, P., Werner, M.~W., \& Zmuidzinas, J., 2001, \apj, 561, 823 

\bibitem[Lis et al. (2002a)]{lgp02} Lis, D.~C., Gerin, M., Phillips, T.~G., \& Motte, F.\ 2002a, \apj, 569, 322 

\bibitem[Lis et al. (2002b)]{lrg02} Lis, D.~C., Roueff, E., Gerin, M., et al., 2002b, \apjl, 571, L55 

\bibitem[Lis et al. (2006)]{lgr06} Lis, D.~C., Gerin, M., Roueff, E., Vastel, C., \& Phillips, T.~G.\ 2006, \apj, 636, 916

\bibitem[Loinard et al. (2002)]{lcc02} Loinard, L., Castets, A., Ceccarelli, C., 2002, P\&SS, 50, 1205

\bibitem[Marcelino et al. (2005)]{mcr05} Marcelino, N., Cernicharo, J., Roueff, E., Gerin, M., \& Mauersberger, R., 2005, \apj, 620, 308 

\bibitem[Maret et al. (2006)]{mbl06} Maret, S., Bergin, E.~A., \& Lada, C.~J., 2006, \nat, 442, 425 

\bibitem[Mathis et al. (1977)]{mrn77} Mathis, J.~S., Rumpl, W., \& Nordsieck, K.~H., 1977, \apj, 217, 425 

\bibitem[McKee (1989)]{m89} McKee, C.~F.\ 1989, \apj, 345, 782

\bibitem[Millar et al. (1989)]{mbh89} Millar, T.~J., Bennett, A., \& Herbst, E., 1989, \apj, 340, 906 

\bibitem[Myers et al. (1991)]{mlf91} Myers, P.~C., Ladd, E.~F., \& Fuller, G.~A., 1991, \apjl, 372, L95 

\bibitem[{\"O}berg et al. (2005)]{ovf05} {\"O}berg, K.~I., van Broekhuizen, F., Fraser, H.~J., et al., 2005, \apjl, 621, L33 

\bibitem[Oliveira et al. (2003)]{ohh03} Oliveira, C.~M., H{\'e}brard, G., Howk, J.~C., et al., 2003, \apj, 587, 235 

\bibitem[Osterbrock (1989)]{ost89} Osterbrock, D.E., 1989, Book Review: Astrophysics of gaseous nebulae and active galactic nuclei. / University Science Books, Astronomy, vol. 17, no. 8, p. 102

\bibitem[Padoan \& Scalo (2005)]{ps05} Padoan, P., \& Scalo, J.\ 2005, \apjl, 624, L97 

\bibitem[Pagani et al. (2003)]{plb03} Pagani, L., Lagache, G., Bacmann, A., 2003, A\&A 406, L59

\bibitem[Pagani et al. (2004)]{pbm04} Pagani, L., Bacmann, A., Motte, F., 2004, A\&A 2004, 417, 605

\bibitem[Pagani et al. (2005)]{ppa05} Pagani, L., Pardo, J.-R., Apponi, A.~J., Bacmann, A., \& Cabrit, S.\ 2005, \aap, 429, 181

\bibitem[Pagani et al. (2007)]{pbc07} Pagani, L., Bacmann, A., Cabrit, S., \& Vastel, C.\ 2007, \aap, 467, 179 

\bibitem[Parise et al. (2004)]{pch04} Parise, B., Castets, A., Herbst, E., et al., 2004, \aap, 416, 159 

\bibitem[Parise et al. (2006)]{pct06} Parise, B., Ceccarelli, C., Tielens, A.~G.~G.~M., et al., 2006, \aap, 453, 949 

\bibitem[Parise et al. (2002)]{pct02} Parise, B., Ceccarelli, C., Tielens, A. G. G. M., et al., 2002, A\&A 393, L49

\bibitem[Pillai et al.(2007)]{pwh07} Pillai, T., Wyrowski, F., Hatchell, J., Gibb, A.~G., \& Thompson, M.~A., 2007, \aap, 467, 207 

\bibitem[Roberts \& Millar (2000a)]{rm00a} Roberts, H., \& Millar, T.~J.\ 2000a, \aap, 364, 780 

\bibitem[Roberts \& Millar (2000b)]{rm00b} Roberts, H., \& Millar, T.~J.\ 2000b, \aap, 361, 388 

\bibitem[Roberts et al. (2003)]{rhm03} Roberts, H., Herbst, E., \& Millar, T.~J.\ 2003, \apjl, 591, L41

\bibitem[Roberts et al. (2004)]{rhm04} Roberts, H., Herbst, E., \& Millar, T.~J.\ 2004, \aap, 424, 905 

\bibitem[Roueff et al. (2005)]{rlv05} Roueff, E., Lis, D.~C., van der Tak, F.~F.~S., Gerin, M., \& Goldsmith, P.~F., 2005, \aap, 438, 585

\bibitem[Schnee \& Goodman (2005)]{sg05} Schnee, S., \& Goodman, A.\ 2005, \apj, 624, 254 

\bibitem[Schnee et al. (2007)]{scg07} Schnee, S., Caselli, P., Goodman, A.~A., et al., 2007, ApJ, 671, 1839

\bibitem[Schnee et al. (2007)]{skg07} Schnee, S., Kauffmann, J., Goodman, A., \& Bertoldi, F.\ 2007, \apj, 657, 838 

\bibitem[Shu et al. (1987)]{sal87} Shu, F.~H., Adams, F.~C., \& Lizano, S., 1987, \araa, 25, 23 

\bibitem[Stamatellos et al. (2007)]{sww07} Stamatellos, D., Whitworth, A.~P., \& Ward-Thompson, D., 2007, MNRAS, 379, 1390

\bibitem[Stark et al. (1999)]{svv99} Stark, R., van der Tak, F.~F.~S., \& van Dishoeck, E.~F., 1999, \apjl, 521, L67

\bibitem[Stepnik et 
al.(2003)]{sab03} Stepnik, B.,  Abergel, A., Bernard, J.-P., et al.\ 2003, \aap, 398, 551 

\bibitem[Sundstr\"om et al.(1994)]{smd94} Sundstrom, G., Mowat, J.R., Danared, H., et al., 1994, Science, 263, 785 

\bibitem[Tafalla et al. (2002)]{tmc02} Tafalla, M., Myers, P.~C., Caselli, P., Walmsley, C.~M., \& Comito, C., 2002, \apj, 569, 815 

\bibitem[Tafalla et al. (2004)]{tmc04} Tafalla, M., Myers, P.~C., Caselli, P., \& Walmsley, C.~M., 2004, \aap, 416, 191 

\bibitem[Tafalla et al. (2006)]{tsm06} Tafalla, M., Santiago-Garc{\'{\i}}a, J., Myers, P.~C., et al., 2006, \aap, 455, 577 

\bibitem[Tielens (2005)]{t05} Tielens, A.~G.~G.~M., 2005, The Physics and Chemistry of the Interstellar Medium, Cambridge University Press,  2005.

\bibitem[Tin{\'e} et al. (2000)]{trf00} Tin{\'e}, S., Roueff, E., Falgarone, E., Gerin, M., \& Pineau des For{\^e}ts, G.\ 2000, \aap, 356, 1039

\bibitem[Turner (1990)]{t90} Turner, B.~E.\ 1990, \apjl, 362, L29

\bibitem[Umebayashi \& Nakano (1990)]{un90} Umebayashi, T., \& Nakano, T.\ 1990, \mnras, 243, 103 

\bibitem[van der Tak et al. (2002)]{vsm02} van der Tak, F.~F.~S., Schilke, P., M{\"u}ller, H.~S.~P., et al., 2002, \aap, 388, L53

\bibitem[van der Tak et al. (2005)]{vcc05} van der Tak, F.~F.~S., Caselli, P., \& Ceccarelli, C., 2005, \aap, 439, 195 

\bibitem[van der Tak et al.(2006)]{vbs06} van der Tak, F.~F.~S., Belloche, A., Schilke, P., et al., 2006, \aap, 454, L99 

\bibitem[van der Tak et al. (2007)]{vbs07} van der Tak, F.~F.~S., Black, J.~H., Sch{\"o}ier, F.~L., Jansen, D.~J., \& van Dishoeck, E.~F., 2007, \aap, 468, 627 

\bibitem[Vastel et al. (2000)]{vcc00} Vastel, C., Caux, E., Ceccarelli, C., et al., 2000, \aap, 357, 994 

\bibitem[Vastel et al. (2003)]{vpc03} Vastel, C., Phillips, T.~G., Ceccarelli, C., \& Pearson, J., 2003, \apjl, 593, L97 

\bibitem[Vastel et al. (2004)]{vpy04} Vastel, C., Phillips, T.~G., \& Yoshida, H., 2004, \apjl, 606, L127

\bibitem[Vastel et al. (2006a)]{vcc06} Vastel, C., Caselli, P., Ceccarelli, C., et al., 2006, \apj, 645, 1198 

\bibitem[Vastel et al. (2006b)]{vpc06} Vastel, C., Phillips, T. G., Caselli, P. et al., 2006, Royal Society of London Transactions Series A, vol. 364, Issue 1848, p.3081-3090

\bibitem[Walmsley et al. (2004)]{wfp04} Walmsley, C.~M., Flower, D.~R., \& Pineau des For{\^e}ts, G., 2004, \aap, 418, 1035 

\bibitem[Ward-Thompson et al. (1995)]{wec95} Ward-Thompson, D., Eiroa, C., \& Casali, M.~M.\ 1995, \mnras, 273, L25

\bibitem[Ward-Thompson et al. (1999)]{wma99} Ward-Thompson, D., Motte, F., \& Andr{\'e}, P.\ 1999, \mnras, 305, 143 

\bibitem[Ward-Thompson et al. (2007)]{wac07} Ward-Thompson, D., Andr{\'e}, P., Crutcher, R., et al., 2007, Protostars and Planets V, 33 

\bibitem[Weingartner \& Draine (2001)]{wd01} Weingartner, J.~C., \& Draine, B.~T., 2001, \apj, 548, 296 

\bibitem[Willacy et al. (1998)]{wlv98} Willacy, K., Langer, W.~D., \& Velusamy, T., 1998, \apjl, 507, L171 

\bibitem[Young et al. (2004)]{yle04} Young, K.~E., Lee, J.-E., Evans, N.~J., II, Goldsmith, P.~F., \& Doty, S.~D., 2004, \apj, 614, 252 

\bibitem[Zucconi et al. (2001)]{zwg01} Zucconi, A., Walmsley, C.~M., \& Galli, D., 2001, \aap, 376, 650 
%
\end{thebibliography}
\end{document}